\documentclass[aps,pra,twocolumn]{revtex4-1}
\usepackage{amsmath}
\usepackage{amssymb}
\usepackage{braket}
\usepackage{graphicx}
\usepackage{caption}
\usepackage{subcaption}
\usepackage[dvipsnames]{xcolor}
\usepackage{xcolor}

\usepackage[colorlinks=true,
             linkcolor=magenta,
             citecolor=magenta,
             urlcolor=magenta,
             pdfencoding=auto, psdextra]{hyperref}

\begin{document}

%\linenumbers

\title{Squeezing-Fueled Quantum Otto Engine via Measurement-Induced Cooling: The Two-Qubit Quantum Rabi Model}

\author{S. R. Rathnakaran}
\email{s.21phz0010@iitrpr.ac.in}
\author{Asoka Biswas}
%\email{abiswas@iitrpr.ac.in}
\affiliation{Department of Physics, Indian Institute of Technology-Ropar, Rupnagar, Punjab, India}

\date{\today}  

\newcommand{\Tr}{\operatorname{Tr}}
\newcommand{\Ha}{\hat{H}_A}
\newcommand{\Hfull}{\hat{H}}
\newcommand{\saz}{\hat{\sigma}_z^A}
\newcommand{\Up}{\ket{++}}
\newcommand{\Down}{\ket{--}}
\newcommand{\rhoG}{\hat{\rho}^{\mathrm{global}}}
\newcommand{\rhoQ}{\hat{\rho}^{\rm qb}}
\newcommand{\rhoCav}{\hat{\rho}^{\mathrm{cav}}}
\newcommand{\Uop}{\hat{\mathcal{U}}}
\newcommand{\Pop}{\mathcal{P}}
\newcommand{\Phit}{\Phi(t_h)}
\newcommand{\al}{\alpha}
\newcommand{\Sz}{S(r,\phi)}
\newcommand{\Disp}[1]{D\!\left(#1\right)}
\newcommand{\TrCav}{\mathrm{Tr}_{\mathrm{cav}}}
\newcommand{\order}[1]{\mathcal{O}\!\left(#1\right)}

\newcommand{\ad}{\hat{a}^{\dagger}}
\newcommand{\aq}{\hat{a}}
\newcommand{\Sq}{\hat{S}(r,\phi)}
\newcommand{\Dhat}{\hat{D}}
\newcommand{\Hc}{\hat{H}_{c}}
\newcommand{\BL}{B_L}
\newcommand{\BH}{B_H}
\newcommand{\bH}{\beta_H \BH}
\newcommand{\rhoD}{\hat{\rho}_D^{\,\mathrm{qb}}}
\newcommand{\Wsq}{W_{\mathrm{sq}}}
\newcommand{\Wm}{W_{\mathrm{meas}}}
\newcommand{\etaO}{\eta_{\mathrm{Otto}}}
\newcommand{\etat}{\eta_{\mathrm{tot}}}

\begin{abstract}
We investigate a quantum Otto engine (QOE) constructed from the two-qubit quantum Rabi model, operating within a cavity quantum electrodynamics (QED) architecture. The engine operates with two qubits as the working substance and a single non-Markovian hot thermal bath, modeled via the hierarchical equations of motion (HEOM) formalism. In place of a conventional cold thermal reservoir, the cooling stroke is realized through a projective measurement protocol on the cavity mode, which acts as an ancillary subsystem and effectively mimics a cold bath for the qubit working medium via measurement back-action. A squeezing drive applied to the cavity mode serves as a quantum fuel. We demonstrate that cavity squeezing systematically enhances both the power output and operational efficiency of the engine - the work extracted per unit of heat drawn from the hot bath-driving it above the standard quantum Otto limit. In the limit-cycle regime, the efficiency, while remaining above the Otto bound throughout, asymptotically converges to it from above. This identifies squeezing as a controllable quantum resource for thermodynamic optimization. Our results reveal that the interplay between qubit-cavity coupling, measurement-induced cooling, and non-equilibrium squeezing gives rise to a multi-resource thermodynamic architecture with performance characteristics inaccessible to conventional two-bath quantum Otto engines, thereby providing a concrete route toward experimentally realizable quantum heat engines in cavity QED platforms.
\end{abstract}
\maketitle

\section{Introduction}

Quantum thermodynamics seeks to understand how work, heat, and efficiency emerge from quantum dynamics and whether quantum resources, namely, coherence, entanglement, squeezing, and measurement back-action, can fundamentally modify energy exchange processes~\cite{Vinjanampathy,deffner2019quantum,science.1078955,hardal2015superradiant,robnagel,alicki1979quantum,kosloff2014quantum}. Quantum heat engines, and the quantum Otto cycle in particular, provide a central framework for probing finite-time and strong-coupling thermodynamics~\cite{QuanHT,Abah}. Cavity quantum electrodynamics (QED) offers an ideal experimental platform for such engines, enabling precise control over cavity-qubit interactions~\cite{haroche2006exploring}; in particular, two-qubit quantum Rabi models (QRMs) operating in the ultrastrong coupling regime realize working media built from collective cavity-qubit states rather than bare atomic excitations~\cite{Barrios}.
 
The two-qubit QRM exhibits rich physics. Grimaudo {\it et al.}~\cite{Grimaudo} showed that it is integrable via a unitary reduction to two decoupled subspaces, revealing first-order quantum phase transitions (QPTs) with abrupt changes in magnetization, photon number, and entanglement~\cite{Forn-D,frisk2019ultrastrong}. Proximity to QPTs amplifies system response~\cite{Fernandez-Lorenzo}, and QPTs in the QRM enhance quantum Stirling engine efficiency toward the Carnot limit~\cite{Wang}, an approach extended to the two-qubit case in~\cite{Xu}. These findings motivate our use of the two-qubit QRM as the working substance for a QOE.
 
A central feature of our work is that cavity squeezing serves as a genuine quantum fuel. Coupling quantum Otto cycles to squeezed reservoirs can push efficiency at maximum power beyond the Carnot limit~\cite{robnagel}, squeezing-induced nonequilibrium states raise the accessible free energy~\cite{Manzano,Varinder,Denzler}, and squeezing enables work extraction from single thermal baths~\cite{Huang,niedenzu2016operation,Yi,PhysRevE.98.042122}. Our engine departs from the conventional two-bath Otto cycle in two additional ways. First, we employ a \emph{single} non-Markovian hot bath modeled via the hierarchical equations of motion (HEOM) formalism~\cite{tanimura1989time,ishizaki2005quantum,Breuer}, capturing non-perturbative system--bath correlations that constitute an exploitable thermodynamic resource~\cite{bhattacharya2020thermodynamic,Ptaszy,Zambon,pezzutto2019out,kamin2020non}. Second, the cold bath is replaced by a projective measurement on the cavity mode. Note that, since Maxwell's era, measurements have enabled work extraction from single-temperature baths~\cite{science.1078955,PhysRevLett.87.220601,Huang,Lin}, and ancilla-mediated protocols~\cite{PhysRevA.108.062214,Yi,PhysRevE.98.042122,Buffoni,PhysRevE.95.032111,elouard2017role,Yi,mohammadi2024quantum,Rathnakaran} cool the working medium via measurement back-action without thermal contact. 
\textcolor{black}{Several forms of measurements, namely, local and entangling measurements ~\cite{Bresque_PhysRevLett.126.120605}, incompatible and generalized measurements~\cite{Manikandan_PhysRevE.105.044137,Lisboa_PhysRevA.106.022436}, and their operational limits and refrigeration analogues \cite{Perna_PhysRevE.109.044102, Elouard_hpgc-nsmr} have been charted. We further note that the thermodynamic cost of quantum measurement is bounded by the acquired information~\cite{Deffner_PhysRevE.94.010103,latune2025thermodynamically}, whose magnitude depends on the nature of the meter~\cite{Linpeng_PhysRevLett.128.220506, Linpeng_PhysRevResearch.6.033045} and, in the idealized projective limit, diverges altogether~\cite{guryanova2020ideal}. Third, we apply squeezing directly to the cavity mode as a controllable quantum fuel, contrary to the conventional studies, in which reservoirs are squeezed to to enhance engine performance beyond conventional thermal bounds~\cite{robnagel, Klaers_PhysRevX.7.031044,Manzano, monika2025asymmetric}. }
Finite-time analysis is essential for revealing quantum effects absent in quasistatic evolution~\cite{PhysRevE.103.032144,PhysRevE.107.054110}, yet the synergy of measurement-based cooling and finite-time dynamics in single-bath cavity-QED engines and their impact on efficiency bounds~\cite{elouard2017role,Buffoni} remains unexplored.

%\textcolor{black}{Second, and inseparable from the first, is .  The present engine unites all these threads-a measurement-based cold stroke and cavity squeezing as fuel---within a single non-Markovian cavity-QED architecture.}
 
Taken together, a two-qubit working substance with cavity squeezing as quantum fuel, a non-Markovian hot bath, and a measurement-based cold stroke, constitute a multi-resource thermodynamic architecture not previously investigated in the literature. We provide a detailed account of how each resource individually and cooperatively enhances QOE performance.
 
The remainder of the paper is organized as follows. In Section~\ref{section2}, we introduce the two-qubit QRM with cavity squeezing that constitutes the working substance and describe the complete quantum Otto cycle in detail. Particularly speaking, we describe how the non-adiabatic expansion and compression strokes are driven by a linear magnetic-field ramp, and how the hot isochoric stroke can be modeled via the hierarchical equations of motion (HEOM) formalism, capturing non-Markovian system-bath correlations. We will also give an outline of how the measurement-induced cold stroke can be realized through projective measurements on displaced cavity Fock states. Within this framework, we develop a perturbative treatment of squeezing and derive analytical expressions for the thermodynamic quantities, namely, work and heat absorbed from the hot bath as functions of the squeezing strength, squeezing phase, and bath parameters. In Section~\ref{section3}, we present the numerical results, including the power-efficiency characteristics as functions of squeezing strength and bath temperature, and thermodynamic contour maps in the squeezing parameter space, revealing the phase-sensitive control of engine performance. We will also provide a multi-cycle convergence analysis that establishes the existence of a stable limit cycle. Finally, in Section~\ref{section4} we summarize our main findings and outline the possible direction of further research. 

\section{Implementation of Quantum Otto engine}
\label{section2}
%\subsection{Model}

We consider a system of two interacting qubits coupled to a cavity as the working medium for the QOE. The total Hamiltonian in natural units ($\hbar = 1$) is given by:
\begin{widetext}
    \begin{equation}
    \hat H = B_1(t) \hat\sigma_z^{(1)} + B_2(t) \hat\sigma_z^{(2)} + \omega \hat a^\dagger \hat a + J_x \hat\sigma_x^{(1)} \hat\sigma_x^{(2)} + J_y \hat\sigma_y^{(1)} \hat\sigma_y^{(2)} + J_z \hat\sigma_z^{(1)} \hat\sigma_z^{(2)} + (g_1 \hat\sigma_z^{(1)} + g_2 \hat\sigma_z^{(2)})(\hat a + \hat a^\dagger),
\end{equation}
\end{widetext}
where $\hat\sigma_{x,y,z}^{(j)}$ are the Pauli matrices for the $j^{th}$ qubit ($j=1,2$), $\hat a$ and $\hat a^\dagger$ are the annihilation and creation operators of the cavity mode with frequency $\omega$, and $B_{1,2}(t)$ denote the qubit energy modulation due to applied magnetic field. The qubits interact via $XYZ$-type spin-spin couplings with coefficients $J_{x,y,z}$. The presence of qubit-qubit interactions enriches the structure of the ground state and leads to first-order QPTs at finite couplings, as shown by Grimaudo {\it et al.} \cite{Grimaudo}. These transitions are characterized by abrupt changes in the magnetization, photon number, and concurrence, and they enable critical control over the thermodynamic properties of the system. This makes the model a powerful platform for engineering quantum thermal machines.

This Hamiltonian represents a generalization of the two-qubit QRM, capturing both cavity-qubit interaction (through the $\sigma_z^{(j)} a$ terms) and direct qubit-qubit interactions. Notably, the spin-spin interactions can serve as effective transverse fields and modify the phase structure of the system. Due to the conserved parity symmetry operator $\hat\Pi = \hat\sigma_z^{(1)} \hat\sigma_z^{(2)} (-1)^{\hat a^\dagger \hat a}$, the Hilbert space decomposes into two invariant subspaces A and B, each representing an effective asymmetric single-qubit Rabi model \cite{Braak}. The corresponding Hamiltonians are
\begin{align}
\hat H_A &= B_+(t) \hat \sigma_z^A + J_-  \hat\sigma_x^A + \omega \hat a^\dagger \hat a + g_+ (\hat a + \hat a^\dagger) \hat\sigma_z^A, \\
\hat H_B &= B_-(t) \hat\sigma_z^B + J_+ \hat\sigma_x^B + \omega \hat a^\dagger \hat a + g_- (\hat a + \hat a^\dagger) \hat\sigma_z^B,
\end{align}
where $B_\pm(t) = B_1(t) \pm B_2(t)$, $J_\pm = J_x \pm J_y$, and $g_\pm = g_1 \pm g_2$. We have identified effective qubits in each subspace, with the respective $z$-components of Pauli matrix, as given by, $\saz = (|++\rangle\langle++| - |--\rangle\langle--|)$ and $\hat\sigma_z^B = |S_0\rangle\langle S_0| - |T_0\rangle\langle T_0|$, where $|S_0\rangle = (|+-\rangle-|-+\rangle)/\sqrt{2}$ and $|T_0\rangle = (|+-\rangle+|-+\rangle)/\sqrt{2}$.   The resulting effective dynamics enables analytical treatment via Braak’s solution to the asymmetric QRM \cite{Braak,Grimaudo}. 

Upon imposing the symmetric constraints, $J_x = J_y = J/2, J_z = 0$, $g_1 = g_2 = g/2$, and $B_1(t) = B_2(t) = B(t)/2$, this configuration allows the system to be described in terms of symmetric and antisymmetric spin subspaces. Defining $\vec{S} = \vec{\sigma}^{(1)} + \vec{\sigma}^{(2)}$, the total spin basis divides the Hilbert space into A and B sectors, that is $\mathcal{H} =\mathcal{H}_A \oplus \mathcal{H}_B$. The Hamiltonian then block-diagonalizes into $\hat H = \hat H_A \oplus \hat H_B$, where:
\begin{equation}
\begin{split}
\hat H_A &= B(t) \hat \sigma_z^A + \omega \hat a^\dagger \hat a + g(\hat a^\dagger + \hat a)\hat\sigma_z^A, \\
\hat H_B &= J \hat\sigma_x^B + \omega \hat a^\dagger \hat a\;.
\end{split}
\end{equation}
Here, $\hat H_A$ describes the subspace consisting of the $|\pm\pm\rangle$ basis, containing the coupled dynamics with the cavity mode, while $\hat H_B$ corresponds to the subspace $(|S_0\rangle,|T_0\rangle)$, which is decoupled from the field and static. These block Hamiltonians form the foundation for analyzing the engine’s dynamics under the Otto cycle. \textcolor{black}{Here the symmetric constraints are imposed to preserve the $\hat\sigma_z^A$ symmetry, $[\hat H_{A}(t),\hat\sigma_z^A]=0$ (and hence the analytic tractability of the cycle), and a time-dependent coupling $g(t)$ (with $g_1=g_2$) constitutes a natural and symmetry-preserving extension. Note that any variation of $J$ does not affect the dynamics of the two qubits in the sector A.}

To this working model, we add a squeezing term:
\begin{equation}
\hat V = \frac{\mu\omega}{2} \left( e^{i\phi} \hat a^2 + e^{-i\phi} \hat a^{\dagger 2} \right)\;,
\label{perturb}
\end{equation}
where $\mu$ is the driving strength (treated as a perturbative parameter) and $\phi$ is the driving phase. This yields the full effective Hamiltonian $\hat H_\text{eff}(t) = \hat H_A(t) + \hat V$, as we focus on the sector A throughout this paper.  This two-qubit Hamiltonian captures the dynamics of strongly coupled qubit and photon, as well as non-classical properties of the cavity mode driven by a squeezing field. It forms the complete model for our QOE analysis. \textcolor{black}{Note that we can confine the dynamics in the sector A only, if we choose an initial condition in the relevant basis $\{|++\rangle,|--\rangle\}$. We did not choose the sector B, because, in this sector, the qubits are effectively decoupled from the magnetic field and the cavity mode, and therefore, it does not support the engine operation and control by quantum resources, e.g., squeezing and measurement.}

\subsection{Quantum Otto Cycle Dynamics}

\begin{figure}
    \centering
    \includegraphics[width=\linewidth]{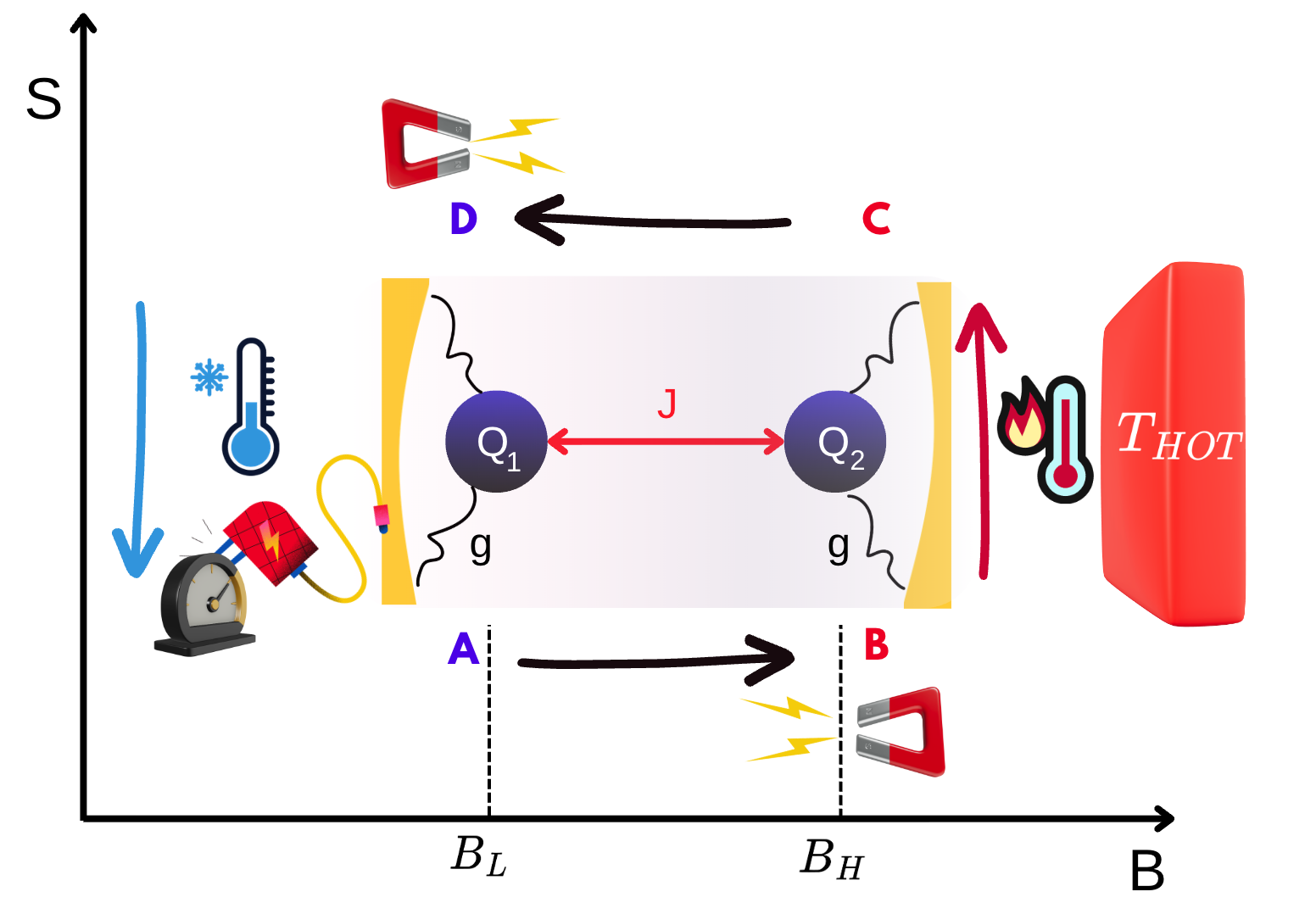}
    \caption{Quantum Otto cycle dynamics comprising of two coupled qubits interacting with a common cavity. }
    \label{fig:QRM1}
\end{figure}

The dynamics of the two–qubit quantum Otto engine proceed through four distinct strokes as shown in Fig.~\ref{fig:QRM1}:

\subsubsection{Expansion Stroke \texorpdfstring{$A \to B$}{A to B}
}
In the first stage, the qubit frequency is non-adiabatically ramped from a lower value $B_L$ to a higher value $B_H$ over a finite time duration $\tau$. The modulation follows a linear schedule: $B(t) = B_L + (B_H - B_L)t/\tau.$
After this transformation, the evolved state of the system is given by $\hat{\rho}_{B}^{\rm global} = \Uop_{\mathrm{exp}}(\tau) \hat\rho_{A}^{\rm global} \Uop^{\dagger}_{\mathrm{exp}}(\tau)$, where $\Uop_{\mathrm{exp}}(\tau) = \mathcal{T}_{\xleftarrow{}} \exp \left[ -i \int_0^{\tau} dt' \, \hat{{H}}_{\text{eff}}(t') \right]$ represents the time evolution operator with explicit time ordering denoted by $\mathcal{T}_{\xleftarrow{}}$. Here $\hat\rho_{A,B}^{\rm global}$ denotes the joint state of the two qubits and the cavity at the beginning and at the end of this stage. 

During a work stroke, the bath is decoupled, and the von Neumann equation gives:
$\dot{\hat{\rho}}^{\rm global} = -i[\hat{H}_{\rm eff}(t), \hat{\rho}^{\rm global}(t)]$
The work done during the stroke $A\to B$ is:
 \begin{equation}
 W_1(\tau)=\int_0^\tau \operatorname{Tr}\left[\hat{\rho}^{\rm qb}\left(t^{\prime}\right) \dot{\hat{H}}_{\rm sys}\left(t^{\prime}\right)\right] d t^{\prime}.
 \label{eq:workAB}
\end{equation}
where $\hat{\rho}^{\rm qb}(t) = \TrCav[\hat{\rho}^{\mathrm{global}}(t)]$ is the reduced qubit density matrix and $\hat{H}_{sys}(t) = B(t)\,\saz$ is the qubit-only Hamiltonian. 

This work can be calculated as $W_1 = \langle E_B^{q}\rangle - \langle E_A^{q}\rangle$, where $ \langle E_A^{q}\rangle$ and $ \langle E_B^{q}\rangle$ indicate the expectation values of the internal energies of the system at the start and at the end of this stage. Note that the energy observables are $\hat{H}_{qi}=B_L\saz$ (at $A$) and
$\hat{H}_{qf}=B_H\saz$ (at $B$). Therefore, the internal energies at the vertices $A$ and $B$ are given as:
\textcolor{black}{
\begin{align}
  \langle E_A^{q}\rangle
    &= \bra{\Psi}B_L\saz\ket{\Psi}
     = -B_L\,\braket{\Psi|\Psi}
     = -B_L,
  \label{eq:EA}\\[6pt]
  \langle E_B^{q}\rangle
    &= \Tr\bigl[B_H\saz\,\rhoQ_B\bigr]
     =  -B_H\;,
  \label{eq:EB}
\end{align}
where $|\Psi\rangle = |--\rangle \otimes |\xi_{--}^{(n)}\rangle$
denotes the initial global state at vertex A of the cycle and $|\xi_{--}^{(n)}\rangle$, as given in Eq.\eqref{eq:init}, is the conditional cavity state in the presence of a perturbation, when the qubits are in $|--\rangle$ state.}

This renders to $\hat{\rho}_A^{\rm global}(0) = |\Psi\rangle\langle \Psi|$. The density matrix $\hat{\rho}_B^{\rm qb}= {\rm Tr}_{\rm cav}(\hat{\rho}_B^{\rm global})$ denotes the state of the two qubits at B.  
Here $\pm1$ is the eigenvalue of $\saz$ on $|\pm \pm\rangle$.
Note that $\langle E_A^{q}\rangle=B_L$ because every component of $\ket{\Psi}$ lies in the $|\pm \pm\rangle$ sector (the squeezing corrections in Appendix \ref{app:A} are all displaced Fock states in the same qubit sector).

So the work done in this stage is calculated as 
\begin{equation}
\textcolor{black}{
W_1 = \langle E_B^{q}\rangle - \langle E_A^{q}\rangle = -(B_H - B_L).}
\label{eq:W1}
\end{equation}

\subsubsection{Hot Isochoric Stroke \texorpdfstring{$B \to C$}{B to C}
}
During the hot isochoric stroke, the qubit frequency is held fixed at $B_H$ while the system interacts with a thermal reservoir at temperature $T$ for a finite time $t_h$, partially thermalising the joint qubit--cavity state. To capture the non-perturbative system--bath correlations beyond the Born--Markov approximation, we employ the hierarchical equations of motion (HEOM) formalism~\cite{Tan}. This formalism provides a numerically exact description of open quantum system dynamics for environments characterized by Gaussian fluctuations and structured spectral densities as given in Appendix~\ref{app:hot_stroke}. 
Since $\hat{H}_{\rm eff}$ is time-independent during this stroke, the heat absorbed by the qubits can be calculated as the energy difference between the final state C and the initial state at B:
\begin{equation}
  Q_H = \int_0^{t_h} \mathrm{Tr}_{\rm sys}\!\left[
    \dot{\hat{\rho}}^{\rm qb}(t')\,\hat{H}_{\rm sys}
  \right] dt'
  = \langle E_C^q \rangle - \langle E_B^q \rangle,
  \label{eq:heatBC}
\end{equation}
where $\hat{H}_{\rm sys} = B_H\saz$.

\textcolor{black}{The qubit working substance couples to its bath through the collective $\hat\sigma_X^A$ operator, which anticommutes with $\hat\sigma_Z^A$ and thereby drives population relaxation. The cavity couples through $(\hat a+\hat a^\dagger)$ and produces only dephasing. We note that $\hat\sigma_X^A$ is a correlated pair-flip operator that preserves the parity of sector $A$, so that relaxation occurs without leakage into sector $B$.}
Consequently, the global state after the hot stroke retains the {\it block-diagonal structure} of the initial state in the subspace $\mathcal{H}_A$:
\begin{widetext}
\begin{equation}
  \rhoG_C = p_{++}(t_h)\,\Up\bra{++}\otimes\ \hat {\Omega}_{++,++}^{\mathrm{cav}}
           + p_{--}(t_h)\,\Down\bra{--}\otimes\hat {\Omega}_{--,--}^{\mathrm{cav}} + \{\zeta(t_h)\,\Up\bra{--}\otimes \hat{\Omega}_{++,--}^{\mathrm{cav}}
           + \mathrm{h.c.}\}.
  \label{eq:rhoC_global}
\end{equation}
\end{widetext}
Here $p_{\pm\pm}(t_h) = \bra{\pm\pm}\hat{\rho}_C^{\rm qb}\ket{\pm\pm}$ (where $\hat{\rho}_C^{\rm qb}=\TrCav\rhoG_C$) are the qubit sector populations satisfying $p_{++}(t_h)+p_{--}(t_h)=1$ and $\Omega_{m,m'}^{\mathrm{cav}}$ $(m,m'\in \{++, --\})$ are cavity operators that encode the qubit-cavity correlations built up during the HEOM evolution.

\paragraph*{Squeezing-induced coherence:}
The off-diagonal element $\zeta(t_h)$ plays a central role in distinguishing the squeezed engine from its unsqueezed counterpart. To understand its origin, we first note that the exact eigenstates of $\hat H_A$ at $r = 0$ are displaced Fock states $\Up \otimes \hat D(-\al)|n\rangle$ and $\Down \otimes \hat D(+\al)|n\rangle$, with displacement amplitude $\al = g/\omega$ (see Appendix~\ref{app:eig analysis}). When squeezing is present, we get the following conditional cavity states for $n=0$:
\begin{equation}
\ket{\xi_{\pm\pm}^{(0)}} = \hat D(\mp\al)\,\hat S(r,\phi)\,\ket{0},
  \label{eq:xi_def}
\end{equation}
where $\hat D(\alpha) = \exp(\alpha \hat a^\dagger - \alpha^* \hat a)$ is the Glauber displacement operator and $\hat S(r,\phi) = \exp[\frac{r}{2}(e^{i\phi}\hat a^2 - e^{-i\phi}(\hat a^\dagger)^2)]$ is the single-mode squeezing operator. Note that the driving strength $\mu$ and the squeezing parameter $r$ are related to each other as $r=2\mu \tau^*$, where $\tau^*$ is the duration of the squeezing field \cite{scully1997quantum}. When $r = 0$, these reduce to the standard coherent states $|\mp\al\rangle$. 

Unlike the coherent states, which have overlap $e^{-2\al^2}$, the displaced-squeezed states $|\xi_{++}^{(0)}\rangle$ and $|\xi_{--}^{(0)}\rangle$ possess a squeezing-modified overlap, as given by 
\begin{equation}
\langle\xi_{++}^{(0)}\ket{\xi_{--}^{(0)}}
    = \exp\!\left(-2\al^2\cosh 2r
                  - 2\al^2\sinh 2r\cos\phi\right)\;.
  \label{eq:overlap_explicit}
\end{equation}

This non-orthogonality has a direct dynamical consequence. Projecting the full HEOM master equation onto the $\Up\bra{--}$ qubit sector and tracing over the cavity (see Appendix~\ref{app:zeta_eom} for details) yields three contributions to $\dot{\zeta}$: (i) free precession at the qubit splitting $2B_H$, (ii) bath-induced dephasing at rate $\Gamma_\phi$ proportional to the $(2\al)^2$ (which is square of the displacement difference of cavity states in two sectors [see Eq. (\ref{eq:xi_def})]), and (iii) a squeezing source term $\mathcal{S}_{\mathrm{sq}}$ that is $\order{r}$ and identically zero when $r = 0$. The resulting equation of motion is
\begin{equation}
  \dot{\zeta}(t)
    = -\bigl(2iB_H + \Gamma_\phi(t)\bigr)\zeta(t)
      + \mathcal{S}_{\mathrm{sq}}(t), 0<t\leq t_h
  \label{eq:zeta_eom}
\end{equation}
where the non-Markovian dephasing rate is $\Gamma_\phi(t) = 4\al^2\int_0^{t}\mathrm{Re}[C_{bath}(t')]\,dt'$ with $C_{bath}(t)$ being the bath correlation function \cite{lambert2023qutipbofin} as given in Appendix \ref{app:hot_stroke}. The squeezing source is $\mathcal{\hat S}_{\mathrm{sq}}(t) = \frac{r\omega}{2}[e^{i\phi}\,\mathcal{M}_{a^2}(t) + e^{-i\phi}\,\mathcal{M}_{(a^\dagger)^2}(t)]$, with $\mathcal{M}_{a^2}(t) = \TrCav[\hat{a}^2\,\Omega_{++,--}^{\mathrm{cav}}(t)]$. When $r = 0$, $\mathcal{S}_{\mathrm{sq}}$ vanishes for all $t$, and the initial condition $\zeta(0) = 0$ (pure state in the $\Up$ sector) ensures $\zeta(t) = 0$ for all $t$. This implies that the unsqueezed HEOM dynamics cannot generate qubit coherence from a diagonal initial state. Squeezing is therefore essential for producing the off-diagonal element, which in turn affects the heat absorbed during this stage.

\paragraph*{Reduced qubit state and populations.}
Tracing Eq.~\eqref{eq:rhoC_global} over the cavity yields the reduced density matrix of the two qubits,
\begin{equation}
  \rhoQ_C = \begin{pmatrix}
               p_{++}(t_h) & \zeta(t_h) \\
               \zeta^*(t_h) & p_{--}(t_h)
             \end{pmatrix},
  \label{eq:rhoC_qb}
\end{equation}
\textcolor{black}{
in the ordered basis $\{\Up,\Down\}$, where the populations can be written as
\begin{align}
  p_{--}(t_h)&=p_{--}^{\rm eq}(t_h)+e^{-\Phi(t_h)}(1-p^{\rm eq}_{--}), \label{eq:p_pp} \\
  p_{++}(t_h) &= {p}_{++}^{\mathrm{eq}}(t_h)\bigl(1-e^{-\Phit}\bigr), \label{eq:p_mm}
\end{align}}
where ${p}_{\pm\pm}^{\rm eq}$ are given by ratios of relative transition rates between the state $|\pm\pm\rangle$ (see Sec. \ref{app:Phi} for details). 

The equilibrium values of these populations are $p_{\pm\pm}^{\mathrm{eq}}(t_h\rightarrow \infty) = e^{-\beta_H(\pm B_H - g^2/\omega)}/Z_H$, where $Z_H = 2\cosh(\beta_H B_H)\,e^{\beta_H g^2/\omega}$ is the partition function. The relaxation kernel $\Phit = \int_0^{t_h}\Gamma_{\mathrm{rel}}(t')\,dt'$ is the time-integrated total relaxation rate, which encodes both the non-Markovian bath memory and the squeezing renormalisation.
In the absence of squeezing, this kernel becomes $\Phi(t_h)=\Gamma_0t_h$, where $\Gamma_0$ is the relaxation rate in the absence of squeezing ($r=0$). In the presence of squeezing, a perturbative expansion in $r$ (detailed in Appendix~\ref{app:Phi}) shows that this damping rate gets renormalized by a term linearly proportional to $r$, as given in Eq. \eqref{eq:B3Geff}.  
\begin{equation}
    \Gamma_{\rm eff}=\Gamma_0\left(1+4r\alpha^2\cos\phi\right)+\mathcal O(r^2)\;.
\end{equation}
The first-order correction includes a squeezing-modified cavity quadrature variance (the associated correction is $r\cos\phi$).
The factor $4\al^2$ in $\Gamma_{\mathrm{eff}}$ is a reminiscence of the squared phase-space separation between the two conditional cavity states, $\hat D(\pm\alpha)|0\rangle$.

Further, the relaxation kernel gets modified by a second-order transient correction arising from virtual photon pairs created by $\hat V$ that persist for a time $\sim 1/\Gamma_{\mathrm{eff}}$ before being damped. As shown through detailed derivation in Appendix~\ref{app:Phi}, the corresponding expression can be written as
given in Eq. \eqref{eq:B3Phi}

\paragraph*{Heat absorbed from the hot bath:}
To compute $Q_H$ [see Eq. (\ref{eq:heatBC})], 
we need to calculate the average energy of the two qubits at C
as given by
\begin{equation}
  \langle E_C^{q}\rangle
    = \Tr\bigl[B_H\saz\,\rhoQ_C\bigr]
    = B_H\,\Pop(t_h).
  \label{eq:EC}
\end{equation}
Here $\Pop(t_h) = p_{++}(t_h) - p_{--}(t_h)$ is the population difference, and can be calculated using Eqs.~\eqref{eq:p_pp}-(\ref{eq:p_mm}) and the relation $p_{++}^{\mathrm{eq}} - p_{--}^{\mathrm{eq}} = -\tanh(\beta_H B_H)$ (see Appendix~\ref{app:Phi}):
\textcolor{black}{
\begin{equation}
  \Pop(t_h)
    = -\tanh(\beta_H B_H)
      - e^{-\Phit}\bigl(1-\tanh(\beta_H B_H)\bigr),
  \label{eq:Pth}
\end{equation}
which interpolates between $\Pop(0)=-1$ (initial pure state) and $\Pop(\infty)=-\tanh(\beta_H B_H)$ (full thermalisation). Note that the coherence $\zeta(t_h)$ does not contribute to the local qubit energy.
Combining with $\langle E_B^{q}\rangle = B_H$ from Eq.~\eqref{eq:EB}, we have
\begin{equation}
  Q_H = B_H\bigl(1-e^{-\Phit}\bigr)\bigl[1-\tanh(\beta_H B_H)  \bigr].
  \label{eq:QH}
\end{equation}
Since $1-\tanh(\beta_H B_H)\ge 0$, we have $Q_H\ge 0$, which indicates that the hot bath injects heat into the system.} The leading-order squeezing correction, obtained by expanding $e^{-\Phit}$ to first order in $r$ is
 
\begin{equation}
  \delta Q_H^{(r)}
    = -B_H\,r\cos\phi\cdot
      \frac{4\al^2\gamma}{\omega}\,
      \bigl(1-e^{-\Gamma_{\mathrm{eff}} t_h}\bigr)
      \bigl[\tanh(\beta_H B_H)-1\bigr].
  \label{eq:dQH}
\end{equation}
% {\bf SHOULD IT BE GAMMA-EFF IN THE ABOVE EXPONENTIAL ?}
We can see that $\delta Q_H^{(r)}\gtrless0 $ for $\cos\phi\gtrless0$, referring to enhanced and reduced heat absorption, respectively. This can be considered as a  \emph{phase-tuneable thermal valve} \cite{PhysRevApplied.10.024003,ronzani2018tunable} - one can change the direction of heat flow, just by changing the phase of the squeezing field.

\subsubsection{Compression Stroke \texorpdfstring{$C \to D$}{C to D}
}
The global state at $C$ is given by Eq. \eqref{eq:rhoC_global}. 
The unitary dynamics during this compression stroke are given by:
\begin{equation}
  \Uop_{\mathrm{com}}(\tau)
    = \Uop_{\mathrm{exp}}(t-\tau).
  \label{eq:Ucom}
\end{equation}

As in the case of expansion stage, since $[H_{\rm eff},\saz]=0$ holds also during the compression stroke, the
unitary $\Uop_{\mathrm{com}}(\tau)$ maps each qubit sector to itself:
\begin{equation}
\begin{split}
\hat{U}_{\mathrm{com}}(\tau) & \bigl( |m\rangle\langle m'| \otimes \Omega_{m,m'}^{\mathrm{cav}} \bigr) \hat{U}_{\mathrm{com}}^\dagger(\tau) \\
&= \bigl[ |m\rangle\langle m'| \otimes \hat{U}_m(\tau) \Omega_{m,m'}^{\mathrm{cav}} \hat{U}_{m'}^\dagger(\tau) \bigr],
\label{eq:sector_pres}
\end{split}
\end{equation}
where $\hat{U}_m(\tau)$ is the effective cavity unitary in sector $m$.
Therefore, the density matrix at D becomes
\begin{equation}
  \rhoQ_D = \TrCav\hat{\rho}_D^{\rm global}
           = \begin{pmatrix}
               p_{++}(t_h) & \zeta_D \\
               \zeta_D^* & p_{--}(t_h)
             \end{pmatrix},
  \label{eq:rhoD_qb}
\end{equation}
where $\hat{\rho}_D^{\rm global}=\Uop_{\mathrm{com}}(\tau)\rhoG_C\Uop_{\mathrm{com}}^\dagger(\tau)$ and the $ \zeta_D = \zeta(t_h)\mathcal{O}_{++,--}(\tau)$ is the coherence modulated by a cross-sector cavity overlap $\mathcal{O}_{++,--}(\tau)$  given by 
\begin{equation}
  \mathcal{O}_{++,--}(\tau)=\TrCav\bigl[\hat{U}_{++}(\tau)\,\Omega_{++,--}^{\mathrm{cav}}\,\hat{U}_{--}^\dagger(\tau)\bigr].
  \label{eq:zeta_D}
\end{equation}
This cross-sector cavity overlap takes the form of a generalised Franck–Condon factor, which quantifies the overlap between the two cavity states that have been displaced to different phase-space positions by the sector-dependent unitaries $\hat{U}_{++}(\tau)$
and $\hat{U}_{--}(\tau)$. This overlap is further multiplied by a squeezing correction $\mathcal{R}_{\mathrm{sq}}(\tau)$ that accounts for the modified quadrature variances, and is given by:
\begin{equation}
  \mathcal{O}_{++,--}(\tau)
    = e^{-|\Delta\al_\tau|^2/2}\,
      e^{i\varphi(\tau)}\,
      \mathcal{R}_{\mathrm{sq}}(\tau),
  \label{eq:overlap_D}
\end{equation}
where $\Delta\al_\tau=\al_{++}(\tau)-\al_{--}(\tau)$ is the difference of
the two displaced amplitudes at the end of the compression stroke. Note that the diagonal populations $p_{++}(t_h)$ and $p_{--}(t_h)$ are \emph{exactly preserved} by the unitary, regardless of $r$, $\tau$, or $g$. 
\textcolor{black}{
The internal energy at $D$  is then given by  %Since $\saz$ isdiagonal in $\{\Up,\Down\}$, the coherence $\zeta_D$ does not contribute:
\begin{eqnarray}
  \langle E_D^{q}\rangle &=&{\rm Tr}(\hat{\rho}_D^{\rm qb}B_L\hat{\sigma}_z^A)\nonumber\\
    &=& B_L\left[-
        \tanh(\beta_H B_H)
        - e^{-\Phit}\bigl\{1-\tanh(\beta_H B_H)\bigr\}
      \right].
  \label{eq:ED_explicit}
\end{eqnarray}
The work done during the stroke $C\to D$ is:
 \begin{equation}
 W_2(\tau)=\int_0^\tau \operatorname{Tr}\left[\hat{\rho}^{\rm qb}\left(t^{\prime}\right) \dot{\hat{H}}_{\rm sys}\left(t^{\prime} - \tau\right)\right] d t^{\prime}.
 \label{eq:workCD}
\end{equation}
Using Eqs.~\eqref{eq:EC} and \eqref{eq:ED_explicit}, the work performed during the $C\to D$ unitary stroke is given by the change in the qubit internal energy, $W_2=\langle E_D^{q}\rangle-\langle E_C^{q}\rangle$,
which explicitly reads
\begin{equation}
W_2=(B_H-B_L)\left[
\tanh(\beta_H B_H)+e^{-\Phi(t_h)}\left(1-\tanh(\beta_H B_H)\right)
\right].
\label{eq:W2}
\end{equation}}

\subsubsection{Cold Stroke via Projective Measurement \texorpdfstring{$D \to A'$}{D to A'}
}
After the unitary return stroke C $\to$ D, the system resides in a global state $\hat{\rho}_D^{\rm global}$. We cool the qubits via a measurement-based stroke that projects the cavity mode onto displaced Fock states. We define the necessary projector as $\Pi_n(\alpha) = D(\alpha)|n\rangle \langle n|D^\dagger(\alpha) \otimes \mathbb{I}_{q_1q_2}$. For each $n$, the cavity is projected, and the post-measurement qubit state is obtained by tracing out the cavity. The conditional qubit state is therefore given by
\begin{equation}
\hat\rho_{D|n}^{\rm qb} = \frac{\mathrm{Tr}_{\mathrm{cav}}(\hat \Pi_n \hat{\rho}_D^{\rm global} \hat \Pi_n)}{\mathrm{Tr}(\hat \Pi_n \hat{\rho}_D^{\rm global})}.
\label{eq:postmeas}
\end{equation}
The probability of each outcome is $p_n = \mathrm{Tr}(\hat\Pi_n \hat\rho_D^{\rm global})$, and the conditional qubit energy is $E_n^{(q)} = \mathrm{Tr}(\hat{H}_q \hat{\rho}_{D|n}^{\rm qb})$, where $\hat{H}_q=B_L\hat{\sigma}_z^A$ is the qubit Hamiltonian.

The maximum cooling is achieved for $n=n_*$, for which the $E_n^{(q)}$ becomes minimum. The energy extracted from the qubit subsystem, therefore, is
\begin{equation}
Q_C = E_{n_*}^{(q)} - \langle E_D^q\rangle.
\end{equation}

This stroke acts as a cold reservoir by probabilistically resetting the qubits into lower-energy states. It harnesses the qubit--cavity correlations induced during the Otto cycle, and displacement $\alpha$ embeds the prior interaction strength $g$. Low-$n$ outcomes correspond to weak cavity excitation and thus to lower qubit energy. From a thermodynamic perspective, this stroke satisfies $Q_C < 0$ and reduces the von Neumann entropy of the qubit subsystem. Hence, it effectively replaces a cold bath in the cycle.

\section{Numerical results}
\label{section3}
\subsection{Power - Efficiency Characteristics under Cavity Squeezing}

\begin{figure}
    \centering
\includegraphics[width=\linewidth]{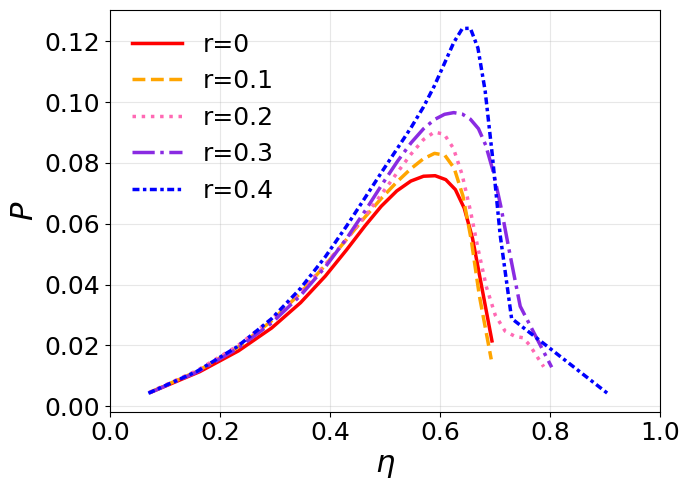}
    \caption{Parametric plot showing the variation of Power (P) vs efficiency ($\eta$) for different squeezing strengths $r$ across the magnetic field varied from $B_L = 0.5$ to $B_H = 10B_L$. The parameters used here are: $\omega_c=5$, $\tau = 1$, $\tau_h = 1$, T = 1, $\alpha = 0.5$, $\phi = 0$. }
    \label{fig:P_vs_eta_r}
\end{figure}

In this section, we present our numerical results. We utilized the HEOM approach in QuTiP \cite{lambert2023qutipbofin}. In Fig.~\ref{fig:P_vs_eta_r} we show that the power-efficiency characteristics of the squeezed two–qubit quantum Otto engine for different values of the squeezing strength $r$, obtained by scanning the magnetic-field ratio $B_H/B_L$ at fixed cavity–qubit coupling $\alpha = 1/2$ and squeezing phase $\phi = 0$. The output power is defined as $P = -W/(2\tau + \tau_h)$, where $W=W_1+W_2$ is the total work per cycle and $2\tau + \tau_h$ is the full cycle duration, while the efficiency is given by $\eta = -W/Q_H$, with $Q_H$ denoting the heat absorbed from the hot bath.

In the absence of squeezing ($r=0$), the engine exhibits a characteristic finite-time Otto behavior: the power increases with efficiency up to an optimal point and then decreases as the cycle approaches the high-efficiency regime. This reflects the competition between enhanced work extraction at larger field ratios and the suppression of power due to reduced population transfer and increased nonadiabatic effects. The maximum power occurs at intermediate efficiencies, well below the Otto bound, indicating the presence of finite-time and coherence-induced losses. As the squeezing strength is increased, a systematic enhancement of the power–efficiency curves is observed. For moderate squeezing ($r=0.1$ and $r=0.2$), both the maximum power and the corresponding efficiency increase compared to the unsqueezed case. This behavior originates from the squeezing-induced modification of the cavity field fluctuations, which effectively renormalizes the qubit–cavity interaction during the compression and expansion strokes. As a consequence, larger work can be extracted per cycle without a proportional increase in cycle duration, leading to an overall enhancement of power. For stronger squeezing ($r \geq 0.3$), the enhancement becomes more pronounced, and the curves extend toward higher efficiencies while maintaining substantial power output. Notably, the peak of the power curve shifts toward larger values of $\eta$, indicating that squeezing allows the engine to operate closer to its optimal thermodynamic regime. This demonstrates that squeezing acts as an additional control parameter that reshapes the power–efficiency trade-off beyond what is achievable by tuning the magnetic-field ratio alone.

\begin{figure*}
    \centering
\includegraphics[width=\linewidth]{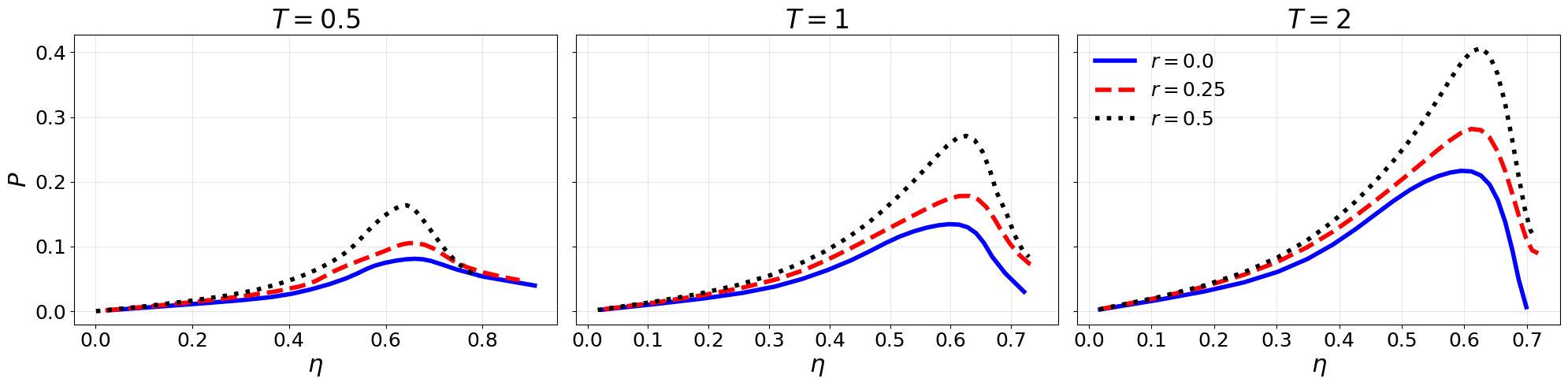}
    \caption{Output power $P$ versus efficiency $\eta$ for different squeezing strengths $r$ at three hot-bath temperatures $T=0.5$, $1$, and $2$. Results are obtained by scanning the magnetic-field ratio $B_H/B_L$ at fixed coupling $\alpha=1/2$ and $\phi=0$, all other parameteric values are the same as in Fig. \ref{fig:P_vs_eta_r}.}
    \label{fig:P_vs_eta_T}
\end{figure*}

The power - efficiency characteristics of the squeezed two–qubit quantum Otto engine for three different hot-bath temperatures, $T=0.5$, $1$, and $2$, at a fixed light–matter coupling $\alpha=1/2$ and squeezing phase $\phi=0$ as shown in Fig.~\ref{fig:P_vs_eta_T}. For each temperature, the output power $P$ is plotted as a function of the efficiency $\eta$ for several values of the squeezing strength $r$, obtained by scanning the magnetic-field ratio $B_H/B_L$. 
The same finite-time Otto trade-off persists at all temperatures, with thermal scaling of the peak power. As the temperature increases, the maximum achievable power increases and the power–efficiency curve extends to higher values of $P$, indicating that a hotter reservoir enhances the overall energy throughput of the engine.

% Please note that squeezing ($r>0$) systematically lifts the power–efficiency curves at all temperatures. For moderate squeezing ($r=0.25$), both the peak power and the efficiency at which it occurs increase compared to the unsqueezed case. This enhancement becomes more pronounced for stronger squeezing ($r=0.5$), where the engine is able to sustain substantially higher power output while operating at larger efficiencies. The effect is particularly striking at higher temperatures, where squeezing and thermal excitations act cooperatively to amplify energy exchange during the unitary strokes.

From a physical perspective, the observed behavior arises from the squeezing-induced modification of the cavity field fluctuations, which effectively renormalizes the qubit–cavity interaction and enhances the level modulation experienced by the qubits during the compression and expansion strokes. At higher temperatures, the hot isochore populates higher-energy eigenstates more efficiently, providing a larger energetic resource that can be partially converted into work. Squeezing further optimizes this conversion by increasing the work extracted per cycle without a proportional increase in cycle duration, thereby boosting the power. At the same time, the location of the maximum power shifts toward lower efficiencies as the temperature increases. This reflects the fact that higher thermal noise enhances dissipation and entropy production, limiting the efficiency at which optimal power can be achieved. Nevertheless, for all temperatures considered, squeezing consistently improves the power–efficiency trade-off, enabling the engine to operate at higher power for a given efficiency compared to the unsqueezed case.

\begin{figure}
    \centering
\includegraphics[width=\linewidth]{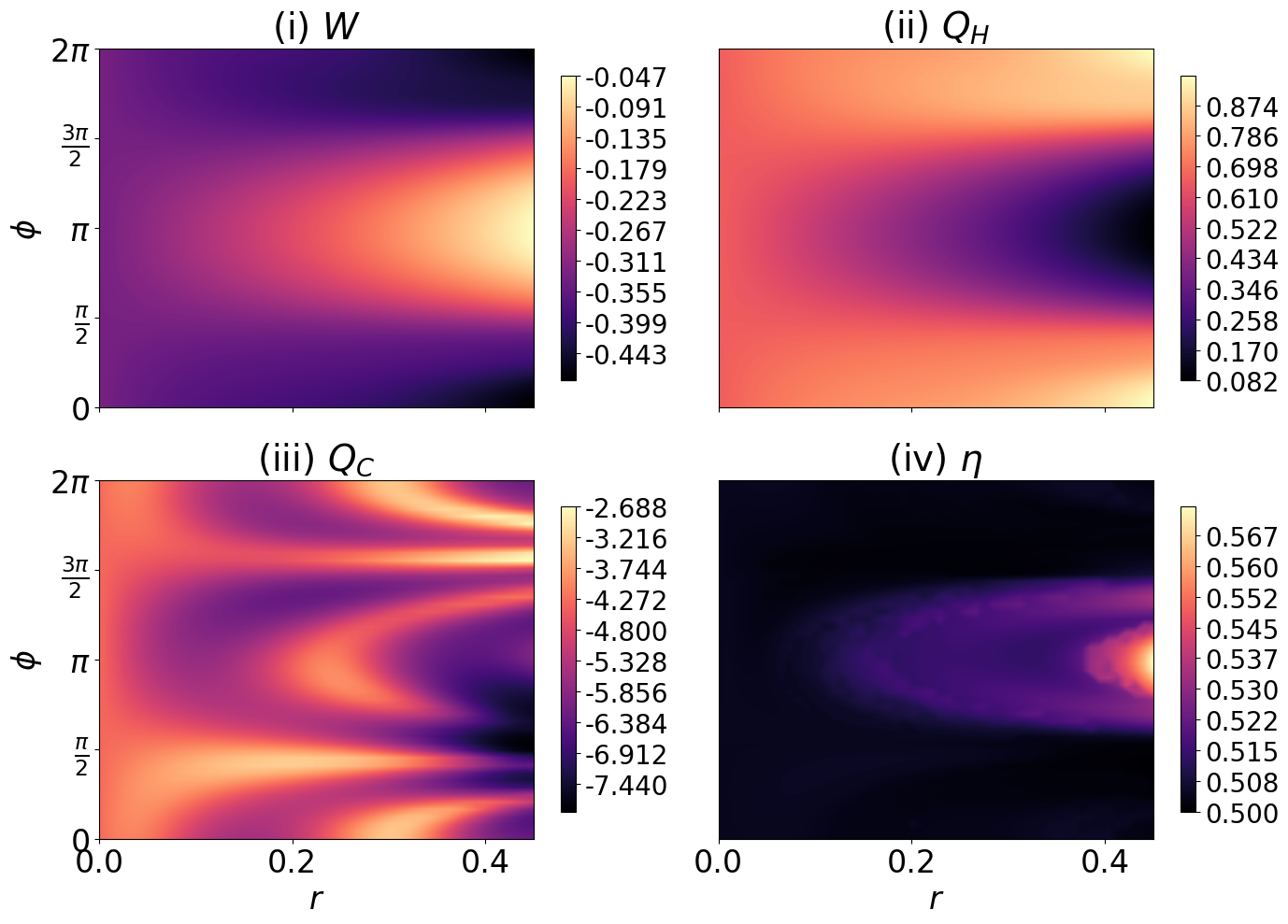}
\caption{Thermodynamic contour maps of the quantum Otto engine as functions of squeezing amplitude r and squeezing phase $\phi$. Shown here (i) total work $W$, (ii) heat input from the hot bath $Q_H$, (iii) cold-stroke energy extraction $Q_C$, and (iv) cycle efficiency $\eta = -W/Q_H$. All parameters are identical to those in Fig.~\ref{fig:P_vs_eta_r}.}
    \label{fig:contour_figures}
\end{figure}

We show in Fig. \ref{fig:contour_figures} the thermodynamic performance of the Otto engine as a function of the squeezing strength $r$ and squeezing phase $\phi$ of the cavity field. The four panels correspond to the net work output $W$, heat absorbed from the hot bath $Q_H$, heat exchanged during the measurement-induced cold stroke $Q_C$, and the resulting efficiency $\eta=-W/Q_H$. These results reveal a pronounced phase-sensitive control of the engine that originates from the anisotropic nature of the squeezed cavity state and its role in the selective measurement process. The extracted work, shown in Fig. \ref{fig:contour_figures}(i), remains negative throughout the explored parameter space, confirming stable engine operation. Its magnitude increases monotonically with the squeezing strength $r$, indicating that stronger squeezing enhances the energetic impact of the measurement stroke. A strong dependence on the squeezing phase is observed, with maximal work extraction occurring near $\phi\simeq\pi$. This behavior can be traced to the alignment of the squeezed quadrature with the cavity operator $(\hat a+\hat a^\dagger)$ that mediates the longitudinal qubit-cavity interaction. At $\phi\simeq\pi$, the measurement becomes optimally sensitive to the qubit energy eigenbasis, leading to a more efficient preparation of low-energy qubit states before the expansion stroke and consequently increased work output. The heat absorbed from the hot reservoir, displayed in Fig. \ref{fig:contour_figures}(ii), is strictly positive for all $(r,\phi)$, as required for engine operation. Notably, $Q_H$ decreases with increasing squeezing strength, particularly in the vicinity of $\phi\simeq\pi$. This reduction reflects the fact that measurement-induced cooling partially replaces the role of the hot bath in establishing population imbalance within the qubit subsystem. As the measurement becomes more energetically selective, a smaller amount of thermal energy is required from the hot bath to sustain the cycle. Fig. \ref{fig:contour_figures}(iii) shows the heat $Q_C$ exchanged during the cold stroke, which is negative across the entire parameter range, indicating energy extraction from the qubits during the measurement process. The magnitude $|Q_C|$ grows with increasing $r$ and exhibits a clear phase dependence mirroring that of the work. Unlike a conventional Otto cycle, the cold stroke here is not mediated by a thermal reservoir but by a selective projective measurement on the cavity field. Post-selection onto the measurement outcome that minimizes the qubit energy effectively implements an information-driven cooling mechanism. The strong phase dependence arises because the squeezing phase controls which cavity quadrature carries information about the qubit energy, thereby determining the efficiency of the measurement-induced projection. The efficiency $\eta$, shown in Fig. \ref{fig:contour_figures}(iv), increases with squeezing strength and attains its maximum near $\phi\simeq\pi$, consistent with the trends observed for $W$, $Q_H$, and $Q_C$. This enhancement originates from the combined effect of increased work extraction and reduced hot-bath heat absorption. 

% Importantly, the efficiency remains well below the Carnot bound for all parameters, ensuring full consistency with the second law of thermodynamics. Although the measurement-assisted cycle benefits from information gain, no violation of thermodynamic principles occurs because the entropy generated in the measurement apparatus and classical record compensates for the apparent reduction in entropy of the qubit subsystem. 
\textcolor{black}{
We note that two distinct coherences appear in the cycle. The cavity is initially prepared in a displaced squeezed state $|\xi^{(0)}_{\pm\pm}\rangle=\hat{D}(\mp\alpha)\hat{S}(r,\phi)|0\rangle$. Its coherence in the Fock basis may be quantified, e.g., by the relative entropy of coherence~\cite{Baumgratz2014}, which grows monotonically with the squeezing strength $r$. %since squeezing progressively broadens the photon-number distribution. 
This cavity also stores a non-passive free energy $\omega\sinh^2 r$ (the ergotropy of the squeezed vacuum; see Appendix~\ref{app:cost}) that fuels the engine, so that the above-Otto efficiency is rooted effectively in the initial cavity coherence. For the qubit, however, squeezing generates the off-diagonal element $\zeta(t_h)$ during the hot stroke [see Eq. \eqref{eq:zeta_eom}], which vanishes identically at $r=0$ and is
therefore a faithful \emph{witness} of the squeezing resource. However, %$\zeta(t_h)$ does not contribute to the qubit energy [see Eq. \eqref{eq:EC}] which depends only on the populations. 
enhancement of the efficiency is mediated by the squeezing-renormalized populations (through $\Gamma_{\mathrm{eff}}$ and the relaxation kernel $\Phi$) and by the non-passive cavity free energy, not by qubit coherence
per se. %Coherence thus serves as the microscopic fingerprint of the squeezing fuel, while the energetics are carried by the population dynamics it accompanies.
}

\subsection{Multi–Cycle Behaviour of the Quantum Otto Engine}

\begin{figure*}
    \centering
\includegraphics[width=\linewidth]{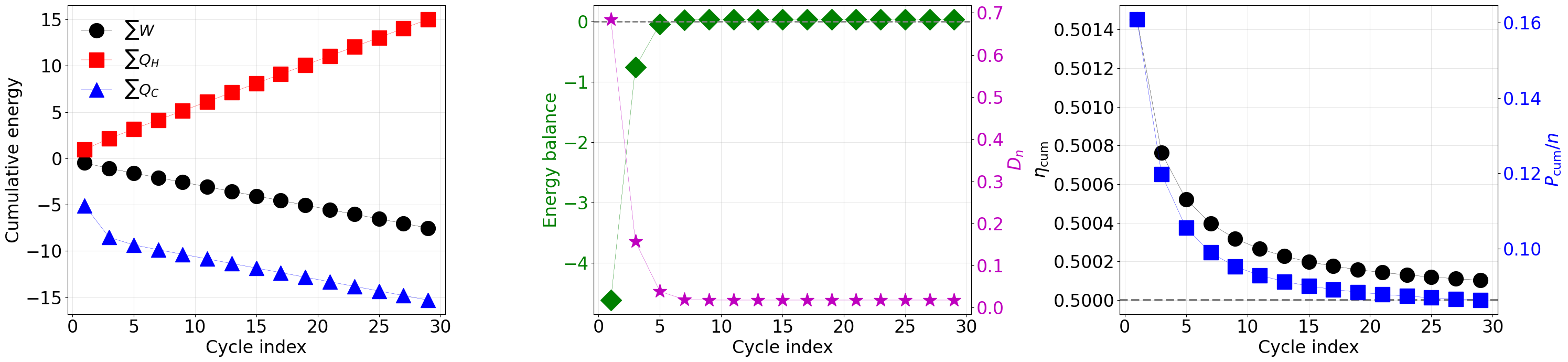}
    \caption{Multi–cycle performance of the squeezed ($r= 0.5$) two–qubit Otto engine. The first panel shows the cumulative work output $\sum_{k\le n} W^{(k)}$, hot heat input $\sum_{k\le n} Q_H^{(k)}$, and cold heat release $\sum_{k\le n} Q_C^{(k)}$ as functions of the cycle index. The middle panel displays the internal energy  $\Delta E_n = W^{(n)}+Q_H^{(n)}+Q_C^{(n)}$ at the end of each cycle together with the trace distance $D_n = \tfrac{1}{2}\|\hat{\rho}^{qb(n)} - \hat{\rho}^{qb(n+1)}\|_1$ between successive initial qubit states. The final panel shows the cumulative efficiency $\eta_{\mathrm{cum}}(n)$ and cumulative power $P_{\mathrm{cum}}(n)$. The parameters used here are $B_L = 0.5$ and $B_H = 2B_L$, and all others are identical to those in Fig. \ref{fig:P_vs_eta_r}.}
    \label{fig:multicycle}
\end{figure*}

In Fig.~\ref{fig:multicycle} we summarize the thermodynamic and dynamical behaviour of the squeezed two–qubit Otto engine operated over $n$ successive cycles. Examining the engine beyond a single cycle is essential, since quantum heat engines generally exhibit non-Markovian transients arising from cavity memory, measurement back-action, and finite-time driving. As a result, the state of the working medium at the beginning of the cycle is not stationary during the first few iterations, and the engine does not immediately operate in a limit cycle. Multi-cycle evolution, therefore, provides direct access to the engine's relaxation dynamics, revealing whether it converges to a steady periodic orbit in which thermodynamic quantities become cycle-invariant and physically meaningful.

The first panel of Fig.~\ref{fig:multicycle} displays the cumulative work, hot heat input, and cold heat dumping as functions of the cycle index. The nearly linear scaling of these accumulated quantities after several cycles demonstrates that the engine reaches a regime in which each cycle contributes approximately the same work and heat. The initial curvature in these cumulative quantities reflects transient memory effects in the cavity–qubit state, but this nonstationarity is rapidly suppressed. The system thus enters a regime in which the cumulative behaviour accurately reflects the engine's long-term energetics. The second panel provides a stability analysis based on two complementary diagnostics: the change in internal energy of the working substance after the cycle $n$: $\Delta E_n = W^{(n)} + Q_H^{(n)} + Q_C^{(n)}$, and the trace distance between the initial qubit states of successive cycles, $D_n = \tfrac{1}{2}\Vert \hat\rho^{{\rm qb}(n)} - \hat\rho^{{\rm qb}(n+1)} \Vert_1$, where $\Vert \cdot \Vert_1$ denotes the trace norm of an operator. The first-law residual $\Delta E_n$ captures the degree to which energy conservation holds on a cycle-by-cycle basis. Its convergence to zero indicates that finite-time strokes and measurement back-action do not cause secular drift in energy balance once the engine has relaxed. Simultaneously, the trace distance quantifies the dynamical convergence of the engine toward a unique limit cycle. The observed rapid decay of $D_n$ confirms that the initial working state becomes cycle invariant: the engine “forgets’’ its starting state and relaxes to a stable orbit in the space of density operators. The combination of these two diagnostics—thermodynamic consistency and dynamical contractivity—establishes that the engine operates in a stable and reproducible manner after the transient regime.

The third panel illustrates the cumulative efficiency and the associated cumulative power, which are given by:
\begin{equation}
\begin{split}
\eta_{\mathrm{cum}}(n) &= \frac{\sum_{k=1}^n W^{(k)}}{\sum_{k=1}^n Q_{H}^{(k)}}\;, \\
P_{\mathrm{cum}}(n) &= \frac{\sum_{k=1}^n W^{(k)}}{n(2\tau + \tau_{\mathrm{h}})}\;.
\end{split}
\end{equation}
Since transient cycles produce atypical contributions to both work and heat exchange, the cumulative quantities offer a physically smoother characterization of the engine’s performance over repeated operation. As the engine approaches its limit cycle, the cumulative efficiency saturates to the Otto bound, $\eta_{\mathrm{Otto}} = 1 - \frac{B_{L}}{B_H}$, demonstrating that the squeezing-enhanced dynamics and the measurement-induced reset do not violate the fundamental thermodynamic structure of the Otto cycle. Concurrently, the cumulative power increases and then stabilizes, reflecting the emergence of a well-defined work output per cycle in the steady regime. The simultaneous saturation of efficiency and stabilization of power confirms that the long-time operation of the engine is both thermodynamically consistent and dynamically stable.

\section{Conclusion and Outlook}
\label{section4}
We have investigated a quantum Otto engine based on a QRM, in which two qubits inside a cavity serve as the working substance. We employ three non-standard thermodynamic ingredients: a single non-Markovian hot bath modeled via the HEOM formalism, 
a projective measurement on the cavity replacing the cold reservoir, and cavity squeezing acting as a quantum fuel. We find analytically tractable expressions for the heat, work, and relaxation kernel through a perturbative treatment of squeezing. 

We find that the cavity squeezing systematically enhances the work delivered per unit of hot-bath heat, the power output, and the efficiency across all bath temperatures and field ratios studied. 
% The efficiency rises strictly above the standard Otto bound $\eta_{\mathrm{Otto}} = 1 - B_L/B_H$, before asymptotically (in many-cycle limit) converging to it from above. 
\textcolor{black}{Such enhancement can be considered as an operational gain that is \emph{supplied and paid for} by the squeezing and measurement resources. However, when accounted for, the work cost of generating the squeezed state and the measurement cost would affect this finding (see Appendix \ref{app:cost}) \cite{Varinder,niedenzu2016operation}. Also, if we were to operate the engine near the quantum critical point (pertaining to the first-order quantum phase transition of the two-qubit Rabi model), where quantum fluctuations and the concurrence are maximal, we could reveal a synergistic enhancement of the efficiency and the power that complements the squeezing fuel demonstrated here.}
%A large value of efficiency is obtained due to the injection of non-passive free energy by the squeezed cavity. 

We further identified a strong dependence on the squeezing phase $\phi$, which acts as a thermodynamic valve. A suitable choice of $\phi$ can tune the heat absorbed from the hot bath, through a relaxation kernel $\Gamma_{\mathrm{eff}}$. We find that maximal work extraction and efficiency occur near $\phi \simeq \pi$, which is attributed to an optimal alignment of the squeezed quadrature with the measurement-sensitive cavity operator. %The non-Markovian bath contributes through a non-trivial squeezing-bath coupling encoded in $\Gamma_{\mathrm{eff}}$, which links the squeezing parameters to the Drude--Lorentz bath cutoff $\gamma$ in a physically transparent manner.
We have also performed a multi-cycle simulation, which confirms a rapid convergence to a steady state value of the efficiency and power. This can be seen from the exponential decay of the trace distance between successive initial qubit states and the vanishing of $\Delta E_n$.

\textcolor{black}{We note that it is indeed possible to operate the device in reverse, by suitably tuning the field ratio and the measurement protocol. This would lead one to a measurement-assisted quantum refrigerator in the ultrastrong coupling regime, in which the squeezing phase would serve as a continuous knob for the coefficient of performance.} 

\section*{Acknowledgement}
We express our sincere gratitude to Dr. Shubhrangshu Dasgupta for his valuable insights and fruitful discussions. S.R.R acknowledges the support from IIT Ropar.

\section*{Data availablity statement}
All data that support the findings of this study are included within the article.

\section*{Conflict of interests}
The authors declare no competing interests.

\appendix
\section{Eigen-spectrum analysis}
\label{app:eig analysis}
\subsection{Exact diagonalization when r = 0}
Block A: This subspace consists of states with total spin $S=1$, and the coupling to the qubits modifies the cavity mode through displacement. The relevant Hamiltonian $\hat H_A$ can be diagonalized with Glauber's displacement operator $\hat D(\alpha) = \exp(\alpha a^\dagger - \alpha^* a)$. The eigenstates and corresponding energies are:
\begin{equation}
|\psi_n^{\pm\pm}\rangle^{(0)} = |\pm\pm\rangle \otimes \hat D(\mp g/\omega)|n\rangle, \quad E_n^{\pm\pm(0)} = \omega n \pm B - \frac{g^2}{\omega}\;.
\label{a1}
\end{equation}

Block B: This subspace corresponds to the independent evolution of the qubits and the cavity states. The eigenstates and corresponding energies are:
\begin{equation}
|\psi_n^{T_0,S_0}\rangle^{(0)} = \frac{1}{\sqrt{2}}(|+-\rangle \pm |-+\rangle) \otimes |n\rangle, \quad E_n^{T_0,S_0} = \omega n \pm J
\end{equation}

As we will see next, the squeezing introduces non-trivial mode mixing, and the relevant first-order corrections to the eigenstates will be written as superpositions of the above zeroth-order eigenstates. \\

\subsection{Perturbative approach for \texorpdfstring{$r \neq 0$}{r neq 0}
}

\subsubsection{Perturbative Energy Corrections in Block A}
\label{app:A}

To understand the influence of squeezing on the eigenenergies of the system, we consider the effective Hamiltonian
\begin{equation}
\hat H_{\rm eff} = \hat H_A + \hat V,
\end{equation}
where $H_A$ denotes the unperturbed Hamiltonian of the two-qubit Rabi model and $V$ is the squeezing term treated as a perturbation, as given in Eq. (\ref{perturb}).
%\begin{equation}
%V = \frac{r\omega}{2}(e^{i\phi}a^2 + e^{-i\phi}(a^\dagger)^2).
%\end{equation}
%Focusing on the symmetric $|\pm \pm\rangle$ sector in Block A, a displacement transformation is applied to diagonalize $H_0$. The unperturbed eigenstates are displaced Fock states, and the energies are:
%\begin{equation}
%E_n^{(0)} = \omega n + B - \frac{g^2}{\omega}.
%\end{equation}
Treating $V$ perturbatively, we compute the energy corrections up to second order. The first-order correction is purely from the expectation value of $V$ in the unperturbed basis $\{D(\mp g/\omega)|n\rangle\}$ of the cavity:
\begin{equation}
E_n^{\pm\pm(1)} = \frac{\mu g^2}{\omega}\cos\phi.
\end{equation}
The second-order corrections involve virtual transitions to neighboring states $|\psi_{n\pm2}^{\pm\pm}\rangle^{(0)}$ and are computed using standard Rayleigh-Schr\"odinger perturbation theory, given by
\begin{equation}
E_n^{\pm\pm(2)} = -\frac{\mu^2\omega}{2}\left(n + \frac{1}{2}\right) - \frac{\mu^2 g^2}{\omega}.
\end{equation}
In block A, the photon states are displaced Fock states due to the qubit-photon interaction. Perturbative corrections to the eigenvectors $|\psi_n^{\pm\pm}\rangle^{(0)}$ arise from contributions of the neighboring displaced states $|\psi_{n\pm1}^{\pm\pm}\rangle^{(0)}$ and $|\psi_{n\pm2}^{\pm\pm}\rangle^{(0)}$.
The resulting first-order correction to the eigenstate is:
\begin{widetext}
\begin{equation}
|\psi_n^{\pm \pm}\rangle^{(1)} = \frac{\mu e^{i\phi} \sqrt{n(n-1)}}{4} |\psi_{n-2}^{\pm\pm}\rangle^{(0)} - {2\mu}{\alpha} e^{i\phi} \sqrt{n} |\psi_{n-1}^{\pm\pm}\rangle^{(0)} \pm \\
 {2\mu}{{\alpha}} e^{-i\phi} \sqrt{n+1} |\psi_{n+1}^{\pm\pm}\rangle^{(0)} + \frac{\mu e^{-i\phi} \sqrt{(n+1)(n+2)}}{4} |\psi_{n+2}^{\pm\pm}\rangle^{(0)}.
 \label{eq:correction}
\end{equation}
\end{widetext}
Here, $\alpha = g/\omega$ is the displacement amplitude. The perturbation induces photon number mixing and reveals the formation of squeezed coherent states.

The complete eigenstate after correction is given by 
\begin{equation}
|\xi^{(n)}_{\pm\pm}\rangle  = |\psi_n^{\pm \pm}\rangle^{(0)} + |\psi_n^{\pm \pm}\rangle^{(1)}
  \label{eq:init}
\end{equation}

For $n=0$, this eigenstate corresponds to the ground state manifold, which can be written as 
\begin{equation}
|\xi^{(0)}_{\pm\pm}\rangle = D(\mp\al)\,\Sz\,|0\rangle.
\label{eq:app_xi_def}
\end{equation}
The overlap between these conditional cavity states is given by \cite{adesso2014continuous}
\begin{align}
\langle\xi^{(0)}_{++}|\xi^{(0)}_{--}\rangle 
&= \langle 0|S^\dagger(r,\phi)\,D^\dagger(-\al)\,D(+\al)\,\Sz|0\rangle\\
&= \langle \mathrm{sq}|\,D(2\al)\,|\mathrm{sq}\rangle,
\end{align}
for real $\alpha$, where $|\mathrm{sq}\rangle \equiv \Sz|0\rangle$.
This yields
\begin{equation}
\langle\xi^{(0)}_{++}|\xi^{(0)}_{--}\rangle 
= \exp\!\Bigl[-2\al^2\bigl(\cosh 2r + \sinh 2r \cos\phi\bigr)\Bigr].
\label{eq:app_overlap_final}
\end{equation}
The overlap is therefore purely real and exhibits exponential suppression governed by both the displacement amplitude $\al$ and the squeezing parameters $r$ and $\phi$. %{\bf DIFFERENT FROM EQ 13} \textcolor{black}{Corrected eqn 13}

\subsubsection{Perturbative Energy Corrections in Block B}

In Block B, the total Hamiltonian is given by $H'_{\rm eff}=H_B+V$. In the absence of the squeezing, the eigenstates correspond to two manifolds, associated with the qubit states  $|T_0\rangle = \frac{1}{\sqrt{2}}(|+-\rangle + |-+\rangle)$, and a singlet $|S_0\rangle = \frac{1}{\sqrt{2}}(|+-\rangle - |-+\rangle)$. In each manifold, the cavity state is that of a harmonic oscillator (not the displaced number state of the block A), without any direct coupling to the qubit subsystem (i.e., $g=0$). %The total Hamiltonian reads:
%\begin{equation}
%H = \omega a^\dagger a + \frac{r\omega}{2}(e^{i\phi} a^2 + e^{-i\phi} (a^\dagger)^2) \pm J,
%\end{equation}
The unperturbed energies are
\begin{equation}
E_n^{T_0,S_0(0)} = \omega n \pm J,
\end{equation}
while the squeezing term introduces no first-order correction
$E_n^{T_0,S_0(1)} = 0$. The second-order corrections yield 
$E_n^{T_0,S_0(2)} = -\frac{\mu^2\omega}{2}\left(n + \frac{1}{2}\right)$,
leading to total energies:
\begin{align}
E_n^{T_0,S_0} &= \omega\left[1 - \frac{\mu^2}{2}\right] n \pm J - \frac{\mu^2\omega}{4}
\end{align}

In the $|T_0\rangle$ and $|S_0\rangle$ sectors, the eigenstates involve standard Fock states (no displacement). Perturbative corrections here involve only two-photon transitions:
\begin{equation}
\begin{split}
|\psi_n^{T_0,S_0}\rangle^{(1)} &= \frac{\mu e^{i\phi} \sqrt{n(n-1)}}{4} |\psi_{n-2}^{T_0,S_0}\rangle^{(0)} \\&+ \frac{\mu e^{-i\phi} \sqrt{(n+1)(n+2)}}{4} |\psi_{n+2}^{T_0,S_0}\rangle^{(0)}.
\end{split}
\end{equation}

\section{Numerical analysis of HEOM}
\label{app:hot_stroke}

Let us start with the Hamiltonian, which describes the interaction of the qubits and the cavity mode with the bath.
\begin{equation}
\hat{H}_{\mathrm{tot}} = \hat{H}_{\mathrm{eff}} + \sum_{\nu}\left(\hat{H}_{\mathrm{E}}^{(\nu)} + \hat{H}_{\mathrm{SE}}^{(\nu)}\right),
\end{equation}
where $\hat{H}_{\mathrm{E}}^{(\nu)}$ describes $\nu$th thermal bath mode, and $\hat{H}_{\mathrm{SE}}^{(\nu)}$denotes the bilinear system-bath coupling, encompassing independent dissipation channel of both the qubits (\textcolor{black}{through collective $\hat\sigma_X^A$ operator}) and the cavity mode (\textcolor{black}{through $(\hat{a} + \hat{a^\dagger)}$}) via their respective bath. Each bath is modeled by a Drude-Lorentz spectral density,
\begin{equation}
J_{\nu}(\omega) = \frac{2 \lambda_{\nu} \gamma_{\nu} \omega}{\omega^{2} + \gamma_{\nu}^{2}},
\label{eq:spec_den}
\end{equation}
with reorganization energy $\lambda_{\nu}$ and bath correlation time $\gamma_\nu^{-1}$. The correlation function of the $\nu$th bath can then be decomposed as $C_{bath}(t) = \sum_{k} c_{\nu k} e^{-\gamma_{\nu k} t}$, which forms the basis of the HEOM hierarchy. Here the decay rate $\gamma_{\nu k}$ of the 
$k$th exponential is given by $\gamma_{\nu 0} = \gamma_\nu$ (which is the Drude cutoff frequency) for the zero-th term, and $\gamma_{\nu k} = \tilde{\nu}_k = 2\pi k/(\beta\hbar)$ for the $k$th Matsubara frequency ($k \geq 1$), $\beta = 1/T$ being the inverse of the temperature $T$ of the bath. 
Also, the $c_{\nu k}$ is the complex amplitude (residue) of the $k$-th exponential in this decomposition, and is given by
$c_{\nu0} = \lambda_\nu \gamma_\nu[\cot(\beta \gamma_\nu/2)-i]$ and
$c_{\nu k} = \frac{4\lambda_\nu\gamma_\nu\tilde{\nu}_k}{\beta(\tilde{\nu}_k^2 - \gamma_\nu^2)}$.

%for the Matsubara terms $k \geq 1$, which are real. 

The reduced dynamics is encoded in auxiliary density operators (ADOs) $\hat{\rho}_{\mathbf{n}}(t)$, with the physical density matrix at the lowest tier $\hat{\rho}_{\mathbf{0}}(t)$. The relevant dynamics are governed by the following equation: 
\begin{widetext}
\begin{align}
\dot{\hat{\rho}}_{\mathbf{n}} = -\left(i \mathcal{L}_{\mathrm{S}} + \sum_{\nu,k} n_{\nu k} \gamma_{\nu k}\right)\hat{\rho}_{\mathbf{n}} \nonumber  - i \sum_{\nu,k} \left[\hat{V}_{\nu}, \hat{\rho}_{\mathbf{n}+\mathbf{e}_{\nu k}}\right]
- i \sum_{\nu,k} n_{\nu k} c_{\nu k} \left(\hat{V}_{\nu} \hat{\rho}_{\mathbf{n}-\mathbf{e}_{\nu k}} - \hat{\rho}_{\mathbf{n}-\mathbf{e}_{\nu k}} \hat{V}_{\nu}\right),
\end{align}
\end{widetext}
where, $\mathcal{L}_{\mathrm{S}}(\cdot) = [\hat{H}_{\mathrm{eff}}, \cdot]$ and $\mathbf{e}_{\nu k}$ is a unit vector in the hierarchy index space. Convergence is verified with respect to hierarchy depth, Matsubara expansion order, and $\mathbf{n}_{\nu k}$. The non-negative integer component $\mathbf{n}_{\nu k}$ of the multi-index $\mathbf{n}$ associated with bath $\nu$
and exponential term $k$ labels the `tier' of the ADO in the hierarchy along the $(\nu, k)$ direction. 
The product $\mathbf{n}_{\nu k}\gamma_{\nu k}$ in the first term acts as a damping coefficient. Since the higher-tier ADOs decay faster, we can truncate the hierarchy at a finite depth. Throughout this paper, we have used the {bath parameters}: {$\gamma_c = 0.1$}, $\gamma_q = 0.1$, $\lambda_c = 0.1$, and $\lambda_q = 0.1$. The \texttt{max depth = 4} and \texttt{Nk = 4}, where \texttt{max depth} is the maximum hierarchy depth to retain, \texttt{Nk} is the number of terms to retain within the expansion of the {Drude-Lorentz Bath}.

\section{Derivation of the coherence equation of motion}
\label{app:zeta_eom}

%\subsubsection{Setup: Projection onto the off-diagonal sector}

The global density matrix in the $\saz$ eigenbasis is given by 
\begin{equation}
\hat{\rho}^{\rm global}(t) = \sum_{m,m' \in \{++,--\}} |m\rangle\langle m'| \otimes \hat{\Omega}_{m,m'}^{\rm cav}(t).
\end{equation}
The off-diagonal coherence is $\zeta(t) = \TrCav[\hat{\Omega}_{++,--}^{\mathrm{cav}}(t)]$. We can obtain its dynamical equation from the Liouville-von Neumann master equation $\dot{\hat{\rho}}^{\rm global} = -i[\hat{H}_{\mathrm{eff}}, \hat{\rho}^{\rm global}] + \mathcal{D}[\hat{\rho}^{\rm global}]$. The part $\hat{H}_A$ in $\hat{H}_{\rm eff}$ leads to a term  $-2iB_H\zeta$ in the equation for $\dot{\zeta}$.  The squeezing perturbation $\hat V = \frac{\mu\omega}{2}(e^{i\phi}\hat a^2 + e^{-i\phi}(\hat a^\dagger)^2)$ generates the following term:
\begin{equation}
\mathcal{S}_{\mathrm{sq}}(t) = \frac{\mu\omega}{2}\bigl[e^{i\phi}\mathcal{M}_{a^2}(t) + e^{-i\phi}\mathcal{M}_{(a^\dagger)^2}(t)\bigr],
\end{equation}
where $\mathcal{M}_{\hat a^2}(t) = \TrCav[\hat{a}^2\,\hat{\Omega}_{++,--}^{\mathrm{cav}}(t)]$. 

In the non-Markovian limit, the dissipator $\mathcal{D}$ for pure dephasing $\saz$ has the following time-integrated form for the dephasing rate:
\begin{equation}
\Gamma_{\phi}(t) 
= 4\alpha^2 \int_0^{t} \mathrm{Re}[C_{\mathrm{bath}}(t')] \, dt'\;,
\end{equation}
where $C_{\rm bath}(t)$ is the Drude-Lorentz bath correlation function \cite{tanimura2020numerically}.

The contributions from the unperturbed Hamiltonian $\hat{H}_A$, the squeezing Hamiltonian $V$, and the dephasing collectively lead to the following equation for the coherence. 
\begin{equation}
\dot{\zeta}(t) = -(2iB_H + \Gamma_\phi)\zeta(t) + \mathcal{S}_{\mathrm{sq}}(t).
\end{equation}

\subsection{Derivation of the relaxation kernel \texorpdfstring{$\Phi(t)$}{Phi(t)} and the effective rate \texorpdfstring{$\Gamma_{\mathrm{eff}}$}{Gammaeff}}
\label{app:Phi}

During the hot isochoric stroke, the populations of the two relevant qubit sectors evolve under bath-induced transitions. Since the population dynamics decouples from coherences in this subspace, it can be written in rate-equation form:
\begin{equation}
\dot p_{++}(t)= -\Gamma_{\downarrow}(t)\,p_{++}(t)+\Gamma_{\uparrow}(t)\,p_{--}(t),
\end{equation}
with $p_{--}(t)=1-p_{++}(t)$. Therefore,
\begin{equation}
\dot p_{++}(t)= -\Gamma_{\rm rel}(t)\,p_{++}(t)+\Gamma_{\uparrow}(t),
\qquad
\Gamma_{\rm rel}(t)\equiv \Gamma_{\downarrow}(t)+\Gamma_{\uparrow}(t).
\label{eq:B3rate}
\end{equation}

The formal solution is
\begin{equation}
p_{++}(t)=p_{++}^{\rm eq}(t)+e^{-\Phi(t)}
\left[p_{++}(0)-p_{++}^{\rm eq}(t)\right],
\end{equation}
where
\begin{equation}
p_{++}^{\rm eq}(t)=\frac{\Gamma_{\uparrow}(t)}{\Gamma_{\rm rel}(t)},
\qquad
\Phi(t)=\int_0^t \Gamma_{\rm rel}(t')dt'.
\end{equation}
At long times, $p_{++}^{\rm eq}(t)$ approaches the thermal stationary population.

The transition rates are determined by the system--bath coupling operator $(\hat a+\hat a^\dagger)$ through matrix elements between the cavity components of the dressed states belonging to the two-qubit sectors:
\begin{equation}
M_{n\to m}=
\!\bra{\psi_m^{--}}
(a+a^\dagger)
\ket{\psi_n^{++}}.
\end{equation}
Within Fermi's golden rule,
\begin{subequations}
\begin{align}
\Gamma_{\downarrow}^{(n\to m)}&\propto |M_{n\to m}|^2J(\omega_{mn})[\bar n(\omega_{mn})+1],\\
\Gamma_{\uparrow}^{(m\to n)}&\propto |M_{n\to m}|^2J(\omega_{mn})\bar n(\omega_{mn}),
\end{align}
\end{subequations}
where $\bar n(\omega)=\frac{1}{e^{\beta\omega}-1}$ is the Bose occupation number. Note that, $\omega_{mn} = E_m^{++} - E_n^{--}$.

For weak coupling, the dominant contribution comes from the lowest transition $n=m=0$. At $r=0$, the lowest cavity states are displaced vacua,
\begin{equation}
\ket{\psi_0^{\pm\pm}}^{(0)}=\ket{\mp\alpha},
\end{equation}
so that
\begin{equation}
|M_{0\to0}|^2=|\bra{\alpha}(a+a^\dagger)\ket{-\alpha}|^2 = 4\alpha^2 e^{-4\alpha^2}.
\end{equation}

Therefore, the unsqueezed downward and upward rates are
\begin{subequations}
\begin{align}
\Gamma_{\downarrow}^{(0 \rightarrow 0)}
&=
4\alpha^2 e^{-4\alpha^2}J(\omega)\,[\bar n(\omega)+1],\\
\Gamma_{\uparrow}^{(0 \rightarrow 0)}
&=
4\alpha^2 e^{-4\alpha^2}J(\omega)\,\bar n(\omega),
\end{align}
\end{subequations}
and the total bare relaxation rate is
\begin{equation}
\Gamma_0=
\Gamma_{\downarrow}^{(0 \rightarrow 0)}+\Gamma_{\uparrow}^{(0 \rightarrow 0)} =
4\alpha^2 e^{-4\alpha^2}J(\omega)\,[2\bar n(\omega)+1].
\label{eq:B3Gamma0}
\end{equation}

When weak squeezing is present ($r\ll1$), the cavity eigenstates become displaced-squeezed states, modifying the Franck-Condon overlap and hence the transition matrix element. Expanding to first order in $r$ gives
\begin{equation}
\frac{|M_{0\to0}^{(r)}|^2}{|M_{0\to0}|^2}
\simeq
1+4r\alpha^2\cos\phi.
\end{equation}
Thus, both upward and downward rates acquire the same multiplicative correction, so that the total effective long-time relaxation rate becomes
\begin{equation}
\Gamma_{\rm eff}=\Gamma_0\left(1+4r\alpha^2\cos\phi\right)+\mathcal O(r^2).
\label{eq:B3Geff}
\end{equation}
Hence, phase-aligned squeezing ($\cos\phi>0$) enhances thermalization, while orthogonal squeezing suppresses it.

We now derive the transient second-order correction. The squeezing perturbation is
\begin{equation}
V=
\frac{\mu\omega}{2}
\left(
e^{-i\phi}\hat a^2+e^{i\phi}\hat a^{\dagger2}
\right),
\end{equation}
which creates or annihilates correlated photon pairs. In the interaction picture of the cavity Hamiltonian $H_c=\omega \hat a^\dagger \hat a$,
\begin{equation}
\hat V(t)=
\frac{\mu\omega}{2}
\left(
e^{-i\phi}\hat a^2e^{-2i\omega t}
+
e^{i\phi}\hat a^{\dagger2}e^{2i\omega t}
\right).
\end{equation}

The leading $r^2$ correction follows from the second-order memory kernel,
\begin{equation}
\delta\Gamma^{(2)}(t)\propto
\int_0^t\,
\langle V(t')V(0)\rangle C_{Bath}(t') \;dt',
\end{equation}
For a Drude-Lorentz bath,
\begin{equation}
C_{bath}(t)\sim e^{-\Gamma_{\rm eff}t},
\end{equation}
% with $\gamma$ {\bf WHERE IS gamma IN B24} the bath bandwidth (see Appendix~\ref{app:hot_stroke}).
After coarse-graining over the fast oscillations at frequency $2\omega$, the remaining envelope gives
\begin{equation}
\delta\Gamma^{(2)}(t)=
-\kappa_0 \mu^2\omega^2 e^{-\Gamma_{\rm eff} t},
\end{equation}
% {\bf SHOULD IT NOT HAVE MU-SQUARE, INSTEAD OF r-SQUARE. SEE B23. PUT B22 IN B23 AND DO THE INTEGRATION EXPLICITLY, TO GET A SUITABLE FORM B25. IT WILL REVEAL AN EXACT EXPRESSION OF ETA. CHECK FROM SCULLY'S BOOK FOR A RELATION BETWEEN MU AND r.}
where $\kappa_0 \approx \al^2 +1/2$ is a dimensionless coefficient containing matrix elements and spectral prefactors after neglecting higher orders.

Accordingly, the full time-dependent relaxation rate becomes
\begin{equation}
\Gamma_{\rm rel}(t)=
\Gamma_{\rm eff}
-\kappa_0 \mu^2\omega^2 e^{-\Gamma_{\rm eff} t}
+\mathcal O(r^3).
\end{equation}

Integrating over the hot-stroke duration $t_h$, we obtain
\begin{align}
\Phi(t_h)&=\int_0^{t_h}\Gamma_{\rm rel}(t')dt'\\
&=
\Gamma_{\rm eff}t_h - \kappa_0\frac{\mu^2\omega^2}{\Gamma_{\rm eff}}
\left(1-e^{-\Gamma_{\rm eff} t_h}\right)
+\mathcal O(r^3).
\label{eq:B3Phi}
\end{align}

The first term is the secular thermal relaxation accumulated during the hot stroke, while the second is a finite-memory correction arising from transient virtual pair fluctuations generated by the squeezing perturbation. For $t_h\gg\Gamma_{\rm eff}^{-1}$, this transient saturates, and the dynamics are governed predominantly by the normalized rate $\Gamma_{\rm eff}$. 
% {\bf IN THE MAIN TEXT, A TIME SCALE 1/GAMMA-EFF WAS MENTIONED. HERE. IT IS gamma-INVERSE. RECHECK.}

\subsection{Compression stroke: density matrix at D}
\label{app:compression}

Since $[\Hfull_{\mathrm{eff}}(t),\saz]=0$ throughout the compression, the unitary admits the block-diagonal decomposition
\begin{equation}
  \Uop_{\mathrm{com}}(\tau)
    = \sum_{m\in\{++,--\}} |m\rangle\langle m|\otimes\hat{U}_m(\tau),
  \label{eq:app_Ublock}
\end{equation}
where $\hat{U}_m(\tau)$ is the effective cavity unitary generated by $H_{\mathrm{eff}}^{(m)}(t)$ in sector $m$.  Applying this to each term of $\rhoG_C$, we have 
\begin{equation}
\Uop_{\mathrm{com}}\bigl(|m\rangle\langle m|\otimes\Omega_{mm}^{\mathrm{cav}}\bigr)\Uop_{\mathrm{com}}^\dagger
  = |m\rangle\langle m|\otimes\hat{U}_m\,\Omega_{mm}^{\mathrm{cav}}\,\hat{U}_m^\dagger, \end{equation}
when $m = m'$ and 
\begin{equation}
   \begin{split}
\Uop_{\mathrm{com}}\bigl(\Up\bra{--}\otimes\Omega_{++,--}^{\mathrm{cav}}\bigr)\Uop_{\mathrm{com}}^\dagger\\
  = \Up\bra{--}\otimes\hat{U}_{++}\,\Omega_{++,--}^{\mathrm{cav}}\,\hat{U}_{--}^\dagger.
\end{split}
\end{equation}
when $m \neq m'$.

The reduced qubit state is $\rhoQ_D = \TrCav[\Uop_{\mathrm{com}}\,\rhoG_C\,\Uop_{\mathrm{com}}^\dagger]$, which can be written in the matrix form as
\begin{equation}
  \rhoQ_D = \begin{pmatrix}
              p_{++}(t_h) & \zeta_D \\[4pt]
              \zeta_D^*   & p_{--}(t_h)
            \end{pmatrix}.
\end{equation}
Note that the off-diagonal element is given by 
\begin{equation}
\zeta_D=\zeta(t_h)\;\TrCav\bigl[{\hat{U}_{++}(\tau)\,\Omega_{++,--}^{\mathrm{cav}}\,\hat{U}_{--}^\dagger(\tau)\bigr]}.
  \label{eq:app_zetaD}
\end{equation}
%For the diagonal elements, the cyclic property of the trace gives
%\begin{equation}
%\begin{split}
%  \bra{m}\rhoQ_D|m\rangle
%    = p_{mm}(t_h)\;\TrCav\bigl[\hat{U}_m\,\Omega_{mm}^{\mathrm{cav}}\,\hat{U}_m^\dagger\bigr]\\
    %= p_{mm}(t_h)\;\TrCav\bigl[\Omega_{mm}^{\mathrm{cav}}\bigr]
%    = p_{mm}(t_h),
%\end{split}
%\end{equation}

%since $\Omega_{mm}^{\mathrm{cav}}$ is a normalised cavity state ($\TrCav[\Omega_{mm}^{\mathrm{cav}}]=1$).  Hence the diagonal populations are \emph{exactly preserved} by the compression, regardless of $r$, $\tau$, or $g$.

%For the off-diagonal element:

During compression, the Hamiltonian in each qubit sector takes the form
\begin{equation}
  H_{\mathrm{eff}}^{(m)}(t)
    = \pm B_{\mathrm{com}}(t) + \omega a^\dagger a \pm g(a+a^\dagger) + V,
\end{equation}
with $B_{\mathrm{com}}(t) = B_H - (B_H-B_L)\,t/\tau$.  The sign of the linear coupling $\pm g(a+a^\dagger)$ shifts the cavity equilibrium to sector-dependent displacements $\al_{\pm\pm}(\tau)$ at the end of the stroke. Now we move to a displaced frame where, 
\begin{equation}
D^\dagger(\alpha_m)\,\hat{a}\,D(\alpha_m) = \hat{a} + \alpha_m= \hat{\tilde{a}}\;.
\end{equation}   
For real $\alpha_m$, we have
\begin{equation}
    \tilde{H}^{(m)}(t) = D^\dagger(\alpha_m)\,H^{(m)}(t)\,D(\alpha_m)\;,
\end{equation}
and the linear coupling $g(\hat{a}+\hat{a}^\dagger)$ cancels exactly, giving rise to the following Hamiltonian:
\begin{equation}
    \tilde{H}^{(m)}(t) = \pm B_{\rm com}(t) + \omega\hat{\tilde{a}}^\dagger\hat{\tilde{a}} - \frac{g^2}{\omega} + \tilde{V}\;.
\end{equation}
The intrinsic unitary is the evolution in this displaced frame, which is given by: 
\begin{equation}
\hat{U}_m^{\rm int}(\tau) = \mathcal{T}_{\xleftarrow{}}\exp\!\left[-i\int_0^\tau \tilde{H}^{(m)}(t')\,dt'\right]
\end{equation}
As $\hat D(\al_m)$ is independent of time, each sector-dependent unitary can therefore be factored as
\begin{equation}
  \hat{U}_m(\tau) = \hat D\!\bigl(\al_m\bigr)\;\hat{U}_m^{\mathrm{int}}(\tau)\;\hat D^\dagger\!\bigl(\al_m\bigr),
  \label{eq:app_Ufactor}
\end{equation}
where $\hat D(\al_m)$ shifts the cavity to the instantaneous equilibrium position, while $\hat U_m^{\mathrm{int}}(\tau)$ describes the remaining harmonic evolution in the displaced frame, including the squeezing perturbation.

Now, substituting Eq.~(\ref{eq:app_Ufactor}) into 
$\mathcal{O}_{++,--}(\tau)=\TrCav[\hat U_{++}\,\Omega_{++,--}^{\mathrm{cav}}\,\hat U_{--}^\dagger]$,
we see that the overlap separates naturally into three contributions: (i) the Franck--Condon factor from the displacement difference $\Delta\al_\tau=\al_{++}(\tau)-\al_{--}(\tau)$, (ii) a relative dynamical phase accumulated between the two sectors, and (iii) a squeezing correction due to the perturbed cavity:
\begin{equation}
\mathcal{R}_{\mathrm{sq}}(\tau)
    \equiv \frac{\TrCav[\hat{U}_{++}^{\mathrm{int}}\hat{\tilde{\Omega}}_{++,--}^{\mathrm{cav}}
         \hat{U}_{--}^{\mathrm{int}\dagger}]}
         {\TrCav[\hat{\tilde{\Omega}}_{++,--}^{\mathrm{cav}}]}.
  \label{eq:app_Rsq_def}
\end{equation}

For $r \neq 0$, a first-order expansion gives:
\begin{equation}
\mathcal{R}_{\mathrm{sq}}(\tau)
    \approx 1 + r\,f(\omega, \phi, \tau) + \mathcal{O}(r^2),
  \label{eq:app_Rsq_expand}
\end{equation}
where $f(\omega, \phi, \tau)$ is a function of the squeezing phase and
the evolution time that is computed numerically from the specific form of
$\hat{U}_m^{\mathrm{int}}$. This gives
\begin{equation}
\mathcal{O}_{++,--}(\tau)= e^{-|\Delta\al_\tau|^2/2}\,
e^{i\varphi(\tau)}\,
\mathcal{R}_{\mathrm{sq}}(\tau),
\end{equation}
where $\mathcal{R}_{\mathrm{sq}}(\tau)\to1$ as $r\to0$, recovering the usual coherent-state overlap.

\section{Thermodynamic cost of the squeezing and measurement resources}
\label{app:cost}
\textcolor{black}{
The efficiency $\eta=-W/Q_H$ reported in the main text exceeds the Otto bound
$\etaO=1-\BL/\BH$. This is an \emph{operational} figure of merit---work output per unit
hot-bath heat---and not a violation of the second law: two non-thermal inputs absent from
$Q_H$ must be accounted for, namely the work $\Wsq$ to prepare the squeezed cavity field and
the cost $\Wm$ of the projective cold stroke~\cite{Manzano,niedenzu2016operation}. We estimate
both below and define a resource-inclusive efficiency $\etat$.
\subsection{Work cost of the squeezing operation}
\label{app:cost:sq}
The squeezed vacuum $|\mathrm{sq}\rangle\equiv\Sq|0\rangle$ is prepared by the operator of
Eq.~(5). The associated Bogoliubov transformation reads
\begin{equation}
\hat S^\dagger \aq\, \hat S = \aq\cosh r - e^{-i\phi}\,\ad\sinh r,
\qquad (\text{h.c. for } \ad), \label{eq:bog}
\end{equation}
whence the mean photon number as follows is phase-independent,
\begin{equation}
\langle \hat n\rangle_{\mathrm{sq}}
=\langle 0|\big(\hat S^\dagger\ad\hat S\big)\big(\hat S^\dagger\aq\hat S\big)|0\rangle
=\sinh^2 r. \label{eq:nsq}
\end{equation}
During preparation the bath is decoupled and the drive acts unitarily, so
$W=\Delta\langle\Hc\rangle$ with $\Hc=\omega\ad\aq$, giving the ideal (reversible) cost
\begin{equation}
\Wsq=\omega\,\langle \hat n\rangle_{\mathrm{sq}}=\omega\sinh^2 r\ \label{eq:Wsq}
\end{equation}
This equals the ergotropy of $|\mathrm{sq}\rangle$. This non-passive free energy constitutes the squeezing
`fuel'~\cite{niedenzu2016operation,Varinder}. For the displaced-squeezed cavity states
$\Dhat(\mp\alpha)\Sq|0\rangle$ used in the cycle, $\langle\hat n\rangle=\alpha^2+\sinh^2 r$, where the first term originates from the displacement $
\alpha$, and the
squeezing-specific cost remains $\omega\sinh^2 r$.
The cost is quadratic, $\Wsq=\omega r^2+O(r^4)$, and must be weighed against the \emph{linear}
work gain. The total work done in one cycle is $W=W_1+W_2$ [see Eqs. (\ref{eq:W1}) and (\ref{eq:W2})]. Expanding $-W$ up to first order in $r$ by Taylor's series, we have $-W = \left.-W\right|_{r=0} + c_1r\cos\phi +\cdots$, where 
%\begin{widetext}
\begin{equation}
c_1=
(\BH-\BL)(1-\tanh\bH)\,e^{-\Gamma_0 t_h}\,4\Gamma_0 t_h\,\alpha^2 >0. \label{eq:gain}
\end{equation}
%\end{widetext}
The net advantage $c_1 r\cos\phi-\omega r^2$ is thus positive over a finite window
$0<r\lesssim r^\ast=c_1\cos\phi/\omega$ ($\cos\phi>0$) and closes at stronger squeezing,
consistent with the saturation discussed in Sec.~III.
\subsection{Information cost of the measurement cold stroke}
\label{app:cost:meas}
The cost of the measurement stroke is fixed by a thermodynamic \emph{bound}, since the projective measurement lowers the entropy of the working substance. % that a cyclic engine must discard. 
For the cold-stroke measurement $\{\hat\Pi_n(\alpha)\}$ [Eq.~\eqref{eq:postmeas}], with conditional qubit states $\hat\rho_{D|n}^{\mathrm{qb}}$ and outcome probabilities $p_n$, the Groenewold information gain~\cite{groenewold1971} is
\begin{equation}
I_G=S\big(\rhoD\big)-\sum_n p_n\,S\big(\hat\rho_{D|n}^{\mathrm{qb}}\big)\;\ge\;0,
\label{eq:groenewold}
\end{equation}
which becomes $S(\rhoD)$ in the ideal limit of selective measurement. The Sagawa-Ueda generalization of the second law~\cite{Sagawa2008} bounds the measurement cost as
$\Wm\;\ge\;T_M\, I_G$, where $T_M$ is the temperature at which the measurement is performed.
In the case of selective measurement, we can calculate this bound with $\rhoD$ of Eq.~\eqref{eq:rhoD_qb}. To the leading order,
\begin{equation}
\Wm\;\gtrsim\;T_M\,H_2\big(p_{++}(t_h)\big)\;,
\label{eq:Wmeas}
\end{equation}
with $H_2(x)=-x\ln x-(1-x)\ln(1-x)$ the binary entropy. This vanishes for a pure qubit and
peaks at $T_M\ln 2$ for a fully mixed qubit ($p_{++}=\tfrac12$). Clearly, ideal projective measurements is unbounded upwards \cite{guryanova2020ideal,latune2025thermodynamically}. }
%Two points: $Q_C$ [Eq.~(32)] is the \emph{energetic} content of the stroke, whereas $\Wm$ bounds the work to \emph{implement and reset} it---distinct ledger entries; and since ideal projective measurements carry an unbounded apparatus cost~, Eq.~\eqref{eq:Wmeas} is the Landauer information floor.
%As we lack an explicit cold bath, we identify $T_M$ with the single reservoir temperature $T$ (the conservative choice); an idealized zero-temperature meter sends $\Wm\to0$.\\

\textcolor{black}{
Considering the minimum measurement cost, the maximum attainable efficiency is therefore given by
\begin{align}
\etat&=\frac{-W}{\,Q_H+\Wsq+\Wm\,} \nonumber \\
&=\frac{-W}{\,Q_H+\omega\sinh^2 r+T_M\,H_2\!\big(p_{++}(t_h)\big)\,} \label{eq:etatot}
\end{align}
%Because $\Wsq,\Wm>0$, one has $\etat<\eta$: once the squeezing and measurement costs enter the denominator, the resource-inclusive efficiency falls below the operational value that exceeds $\etaO$. 
The apparent super-Otto operation is therefore fully consistent with the generalized Carnot bound for a non-equilibrium resource~\cite{Manzano}, and the enhancement is correctly read as a paid-for conversion of non-passive free energy and measurement information into work.}

\twocolumngrid 
\bibliography{main}  % Replace 'references' with your .bib file name

@article{Vinjanampathy,
author = {Sai Vinjanampathy and Janet Anders},
title = {Quantum thermodynamics},
journal = {Contemporary Physics},
volume = {57},
number = {4},
pages = {545--579},
year = {2016},
publisher = {Taylor \& Francis}}

@book{deffner2019quantum,
  title={Quantum Thermodynamics: An introduction to the thermodynamics of quantum information},
  author={Deffner, Sebastian and Campbell, Steve},
  year={2019},
  publisher={Morgan \& Claypool Publishers}
}

@article{hardal2015superradiant,
  title={Superradiant quantum heat engine},
  author={Hardal, AU and M{\"u}stecaplioglu, OE},
  journal={Sci. Rep.},
  volume={5},
  pages={12953},
  year={2015}
}

@article{alicki1979quantum,
  title={The quantum open system as a model of the heat engine},
  author={Alicki, Robert},
  journal={Journal of Physics A},
  volume={12},
  number={5},
  pages={L103--L107},
  year={1979},
  doi={10.1088/0305-4470/12/5/007}
}

@article{kosloff2014quantum,
  title={Quantum heat engines and refrigerators: Continuous devices},
  author={Kosloff, Ronnie and Levy, Amikam},
  journal={Annual Review of Physical Chemistry},
  volume={65},
  number={1},
  pages={365--393},
  year={2014},
  publisher={Annual Reviews},
  doi={https://doi.org/10.1146/annurev-physchem-040513-103724}
}

@book{haroche2006exploring,
  title={Exploring the quantum: atoms, cavities, and photons},
  author={Haroche, Serge and Raimond, J-M},
  year={2006},
  publisher={Oxford University Press}
}

@article{science.1078955,
    author = {Scully, Marlan O.} ,
    title = {Extracting Work from a Single Heat Bath via Vanishing Quantum Coherence},
    journal = {Science},
    volume={299},
    pages={862-864},
    year = {2003}
}

@article{frisk2019ultrastrong,
  title={Ultrastrong coupling between light and matter},
  author={Frisk Kockum, Anton and Miranowicz, Adam and De Liberato, Simone and Savasta, Salvatore and Nori, Franco},
  journal={Nature Reviews Physics},
  volume={1},
  number={1},
  pages={19--40},
  year={2019},
  publisher={Nature Publishing Group UK London}
}

@article{bhattacharya2020thermodynamic,
  title={Thermodynamic utility of non-Markovianity from the perspective of resource interconversion},
  author={Bhattacharya, Samyadeb and Bhattacharya, Bihalan and Majumdar, AS},
  journal={Journal of Physics A},
  volume={53},
  number={33},
  pages={335301},
  year={2020},
  publisher={IOP Publishing}
}

@article{robnagel,
  title = {Nanoscale Heat Engine Beyond the Carnot Limit},
  author = {Ro\ss{}nagel, J. and Abah, O. and Schmidt-Kaler, F. and Singer, K. and Lutz, E.},
  journal = {Phys. Rev. Lett.},
  volume = {112},
  issue = {3},
  pages = {030602},
  numpages = {5},
  year = {2014},
  month = {Jan},
  publisher = {American Physical Society},
  doi = {10.1103/PhysRevLett.112.030602},
  url = {https://link.aps.org/doi/10.1103/PhysRevLett.112.030602}
}

@article{Barrios,
  title = {Role of quantum correlations in light-matter quantum heat engines},
  author = {Barrios, G. Alvarado and Albarr\'an-Arriagada, F. and C\'ardenas-L\'opez, F. A. and Romero, G. and Retamal, J. C.},
  journal = {Phys. Rev. A},
  volume = {96},
  issue = {5},
  pages = {052119},
  numpages = {9},
  year = {2017},
  month = {Nov},
  publisher = {American Physical Society},
  doi = {10.1103/PhysRevA.96.052119},
  url = {https://link.aps.org/doi/10.1103/PhysRevA.96.052119}
}

@article{pezzutto2019out,
  title={An out-of-equilibrium non-Markovian quantum heat engine},
  author={Pezzutto, Marco and Paternostro, Mauro and Omar, Yasser},
  journal={Quantum Science and Technology},
  volume={4},
  number={2},
  pages={025002},
  year={2019},
  publisher={IOP Publishing}
}

@article{QuanHT,
  title = {Quantum thermodynamic cycles and quantum heat engines},
  author = {Quan, H. T. and Liu, Yu-xi and Sun, C. P. and Nori, Franco},
  journal = {Phys. Rev. E},
  volume = {76},
  issue = {3},
  pages = {031105},
  numpages = {18},
  year = {2007},
  month = {Sep},
  publisher = {American Physical Society},
  doi = {10.1103/PhysRevE.76.031105},
  url = {https://link.aps.org/doi/10.1103/PhysRevE.76.031105}
}

@article{Abah,
  title = {Single-Ion Heat Engine at Maximum Power},
  author = {Abah, O. and Ro\ss{}nagel, J. and Jacob, G. and Deffner, S. and Schmidt-Kaler, F. and Singer, K. and Lutz, E.},
  journal = {Phys. Rev. Lett.},
  volume = {109},
  issue = {20},
  pages = {203006},
  numpages = {6},
  year = {2012},
  month = {Nov},
  publisher = {American Physical Society},
  doi = {10.1103/PhysRevLett.109.203006},
  url = {https://link.aps.org/doi/10.1103/PhysRevLett.109.203006}
}

@article{Braak,
  title = {Integrability of the Rabi Model},
  author = {Braak, D.},
  journal = {Phys. Rev. Lett.},
  volume = {107},
  issue = {10},
  pages = {100401},
  numpages = {4},
  year = {2011},
  month = {Aug},
  publisher = {American Physical Society},
  doi = {10.1103/PhysRevLett.107.100401},
  url = {https://link.aps.org/doi/10.1103/PhysRevLett.107.100401}
}

@article{Grimaudo,
  title = {Quantum Phase Transitions for an Integrable Quantum Rabi-like Model with Two Interacting Qubits},
  author = {Grimaudo, R. and de Castro, A. S. Magalh\tilde{a}es and Messina, A. and Solano, E. and Valenti, D.},
  journal = {Phys. Rev. Lett.},
  volume = {130},
  issue = {4},
  pages = {043602},
  numpages = {6},
  year = {2023},
  month = {Jan},
  publisher = {American Physical Society},
  doi = {10.1103/PhysRevLett.130.043602},
  url = {https://link.aps.org/doi/10.1103/PhysRevLett.130.043602}
}

@book{scully1997quantum,
  title={Quantum Optics},
  author={Scully, Marlan O and Zubairy, M Suhail},
  year={1997},
  publisher={Cambridge University Press}
}

@article{Forn-D,
  title = {Ultrastrong coupling regimes of light-matter interaction},
  author = {Forn-D\'{\i}az, P. and Lamata, L. and Rico, E. and Kono, J. and Solano, E.},
  journal = {Rev. Mod. Phys.},
  volume = {91},
  issue = {2},
  pages = {025005},
  numpages = {48},
  year = {2019},
  month = {Jun},
  publisher = {American Physical Society},
  doi = {10.1103/RevModPhys.91.025005},
  url = {https://link.aps.org/doi/10.1103/RevModPhys.91.025005}
}

@article{Klaers_PhysRevX.7.031044,
  title = {Squeezed Thermal Reservoirs as a Resource for a Nanomechanical Engine beyond the Carnot Limit},
  author = {Klaers, Jan and Faelt, Stefan and Imamoglu, Atac and Togan, Emre},
  journal = {Phys. Rev. X},
  volume = {7},
  issue = {3},
  pages = {031044},
  numpages = {6},
  year = {2017},
  month = {Sep},
  publisher = {American Physical Society},
  doi = {10.1103/PhysRevX.7.031044},
  url = {https://link.aps.org/doi/10.1103/PhysRevX.7.031044}
}

@article{monika2025asymmetric,
  title={Asymmetric quantum harmonic Otto engine under hot squeezed thermal reservoir},
  author={Monika and Kaur, Kirandeep and Singh, Varinder and Rebari, Shishram},
  journal={Journal of Non-Equilibrium Thermodynamics},
  volume={50},
  number={3},
  pages={435--445},
  year={2025},
  publisher={De Gruyter}
}

@article{niedenzu2016operation,
  title={On the operation of machines powered by quantum non-thermal baths},
  author={Niedenzu, Wolfgang and Gelbwaser-Klimovsky, David and Kofman, Abraham G and Kurizki, Gershon},
  journal={New Journal of Physics},
  volume={18},
  number={8},
  pages={083012},
  year={2016},
  publisher={IOP Publishing}
}

@article{Fernandez-Lorenzo,
  title = {Quantum sensing close to a dissipative phase transition: Symmetry breaking and criticality as metrological resources},
  author = {Fern\'andez-Lorenzo, Samuel and Porras, Diego},
  journal = {Phys. Rev. A},
  volume = {96},
  issue = {1},
  pages = {013817},
  numpages = {10},
  year = {2017},
  month = {Jul},
  publisher = {American Physical Society},
  doi = {10.1103/PhysRevA.96.013817},
  url = {https://link.aps.org/doi/10.1103/PhysRevA.96.013817}
}

@article{kamin2020non,
  title={Non-Markovian effects on charging and self-discharging process of quantum batteries},
  author={Kamin, FH and Tabesh, Fatemeh T and Salimi, Shahriar and Kheirandish, Fardin and Santos, Alan C},
  journal={New Journal of Physics},
  volume={22},
  number={8},
  pages={083007},
  year={2020},
  publisher={IOP Publishing}
}

@article{tanimura1989time,
  title={Time evolution of a quantum system in contact with a nearly Gaussian-Markoffian noise bath},
  author={Tanimura, Yoshitaka and Kubo, Ryogo},
  journal={Journal of the Physical Society of Japan},
  volume={58},
  number={1},
  pages={101--114},
  year={1989},
  publisher={The Physical Society of Japan}
}

@article{tanimura2020numerically,
  title={Numerically “exact” approach to open quantum dynamics: The hierarchical equations of motion (HEOM)},
  author={Tanimura, Yoshitaka},
  journal={J. Chem. Phys.},
  volume={153},
  number={2},
  pages={020901},
  year={2020},
  publisher={AIP Publishing}
}

@article{ishizaki2005quantum,
  title={Quantum dynamics of system strongly coupled to low-temperature colored noise bath: Reduced hierarchy equations approach},
  author={Ishizaki, Akihito and Tanimura, Yoshitaka},
  journal={Journal of the Physical Society of Japan},
  volume={74},
  number={12},
  pages={3131--3134},
  year={2005},
  publisher={The Physical Society of Japan}
}

@article{elouard2017role,
  title={The role of quantum measurement in stochastic thermodynamics},
  author={Elouard, Cyril and Herrera-Mart{\'\i}, David A and Clusel, Maxime and Auff{\`e}ves, Alexia},
  journal={npj Quantum Information},
  volume={3},
  number={1},
  pages={9},
  year={2017},
  publisher={Nature Publishing Group UK London}
}

@article{mohammadi2024quantum,
  title={Quantum non-Markovianity, quantum coherence and extractable work in a general quantum process},
  author={Mohammadi, Amin and Shafiee, Afshin},
  journal={Physical Chemistry Chemical Physics},
  volume={26},
  number={5},
  pages={3990--3999},
  year={2024},
  publisher={Royal Society of Chemistry}
}

@article{Wang,
  title = {Critical behavior of the quantum Stirling heat engine},
  author = {Wang, Yuan-Sheng and Yung, Man-Hong and Xu, Dazhi and Liu, Maoxin and Chen, Xiaosong},
  journal = {Phys. Rev. A},
  volume = {109},
  issue = {2},
  pages = {022208},
  numpages = {10},
  year = {2024},
  month = {Feb},
  publisher = {American Physical Society},
  doi = {10.1103/PhysRevA.109.022208},
  url = {https://link.aps.org/doi/10.1103/PhysRevA.109.022208}
}

@article{Xu,
  title = {Quantum Stirling heat engine based on two-qubit quantum Rabi model with spin-spin coupling},
  author = {Xu, Luxin and Wu, Chunfeng and Ren, Changliang},
  journal = {Phys. Rev. A},
  volume = {112},
  issue = {3},
  pages = {032226},
  numpages = {13},
  year = {2025},
  month = {Sep},
  publisher = {American Physical Society},
  doi = {10.1103/zqbv-j3jv},
  url = {https://link.aps.org/doi/10.1103/zqbv-j3jv}
}

@article{Huang,
  title = {Effects of reservoir squeezing on quantum systems and work extraction},
  author = {Huang, X. L. and Wang, Tao and Yi, X. X.},
  journal = {Phys. Rev. E},
  volume = {86},
  issue = {5},
  pages = {051105},
  numpages = {6},
  year = {2012},
  month = {Nov},
  publisher = {American Physical Society},
  doi = {10.1103/PhysRevE.86.051105},
  url = {https://link.aps.org/doi/10.1103/PhysRevE.86.051105}
}

@article{Manzano,
  title = {Entropy production and thermodynamic power of the squeezed thermal reservoir},
  author = {Manzano, Gonzalo and Galve, Fernando and Zambrini, Roberta and Parrondo, Juan M. R.},
  journal = {Phys. Rev. E},
  volume = {93},
  issue = {5},
  pages = {052120},
  numpages = {10},
  year = {2016},
  month = {May},
  publisher = {American Physical Society},
  doi = {10.1103/PhysRevE.93.052120},
  url = {https://link.aps.org/doi/10.1103/PhysRevE.93.052120}
}

@article{Varinder,
  title = {Performance bounds of nonadiabatic quantum harmonic Otto engine and refrigerator under a squeezed thermal reservoir},
  author = {Singh, Varinder and M\"ustecapl\ifmmode \imath \else \i \fi{}o\ifmmode \breve{g}\else \u{g}\fi{}lu, \"Ozg\"ur E.},
  journal = {Phys. Rev. E},
  volume = {102},
  issue = {6},
  pages = {062123},
  numpages = {8},
  year = {2020},
  month = {Dec},
  publisher = {American Physical Society},
  doi = {10.1103/PhysRevE.102.062123},
  url = {https://link.aps.org/doi/10.1103/PhysRevE.102.062123}
}

@article{Denzler,
  title = {Power fluctuations in a finite-time quantum Carnot engine},
  author = {Denzler, Tobias and Lutz, Eric},
  journal = {Phys. Rev. Res.},
  volume = {3},
  issue = {3},
  pages = {L032041},
  numpages = {6},
  year = {2021},
  month = {Aug},
  publisher = {American Physical Society},
  doi = {10.1103/PhysRevResearch.3.L032041},
  url = {https://link.aps.org/doi/10.1103/PhysRevResearch.3.L032041}
}

@article{Breuer,
  title = {Colloquium: Non-Markovian dynamics in open quantum systems},
  author = {Breuer, Heinz-Peter and Laine, Elsi-Mari and Piilo, Jyrki and Vacchini, Bassano},
  journal = {Rev. Mod. Phys.},
  volume = {88},
  issue = {2},
  pages = {021002},
  numpages = {24},
  year = {2016},
  month = {Apr},
  publisher = {American Physical Society},
  doi = {10.1103/RevModPhys.88.021002},
  url = {https://link.aps.org/doi/10.1103/RevModPhys.88.021002}
}

@article{Ptaszy,
  title = {Non-Markovian thermal operations boosting the performance of quantum heat engines},
  author = {Ptaszy\ifmmode \acute{n}\else \'{n}\fi{}ski, Krzysztof},
  journal = {Phys. Rev. E},
  volume = {106},
  issue = {1},
  pages = {014114},
  numpages = {11},
  year = {2022},
  month = {Jul},
  publisher = {American Physical Society},
  doi = {10.1103/PhysRevE.106.014114},
  url = {https://link.aps.org/doi/10.1103/PhysRevE.106.014114}
}

@article{Zambon,
  title = {Quantum Processes as Thermodynamic Resources: The Role of Non-Markovianity},
  author = {Zambon, Guilherme and Adesso, Gerardo},
  journal = {Phys. Rev. Lett.},
  volume = {134},
  issue = {20},
  pages = {200401},
  numpages = {9},
  year = {2025},
  month = {May},
  publisher = {American Physical Society},
  doi = {10.1103/PhysRevLett.134.200401},
  url = {https://link.aps.org/doi/10.1103/PhysRevLett.134.200401}
}

@article{Buffoni,
  title = {Quantum Measurement Cooling},
  author = {Buffoni, Lorenzo and Solfanelli, Andrea and Verrucchi, Paola and Cuccoli, Alessandro and Campisi, Michele},
  journal = {Phys. Rev. Lett.},
  volume = {122},
  issue = {7},
  pages = {070603},
  numpages = {5},
  year = {2019},
  month = {Feb},
  publisher = {American Physical Society},
  doi = {10.1103/PhysRevLett.122.070603},
  url = {https://link.aps.org/doi/10.1103/PhysRevLett.122.070603}
}

@article{Yi,
  title = {Single-temperature quantum engine without feedback control},
  author = {Yi, Juyeon and Talkner, Peter and Kim, Yong Woon},
  journal = {Phys. Rev. E},
  volume = {96},
  issue = {2},
  pages = {022108},
  numpages = {5},
  year = {2017},
  month = {Aug},
  publisher = {American Physical Society},
  doi = {10.1103/PhysRevE.96.022108},
  url = {https://link.aps.org/doi/10.1103/PhysRevE.96.022108}
}

@article{Rathnakaran,
  title = {Ancilla measurement-based quantum Otto engine using double-pair spin architecture},
  author = {Rathnakaran, S. R. and Biswas, Asoka},
  journal = {Phys. Rev. E},
  volume = {111},
  issue = {6},
  pages = {064116},
  numpages = {12},
  year = {2025},
  month = {Jun},
  publisher = {American Physical Society},
  doi = {10.1103/w4j5-gftl},
  url = {https://link.aps.org/doi/10.1103/w4j5-gftl}
}

@article{PhysRevLett.87.220601,
  title = {Extracting Work from a Single Thermal Bath via Quantum Negentropy},
  author = {Scully, Marlan O.},
  journal = {Phys. Rev. Lett.},
  volume = {87},
  issue = {22},
  pages = {220601},
  numpages = {4},
  year = {2001},
  month = {Nov},
  publisher = {American Physical Society},
  doi = {10.1103/PhysRevLett.87.220601},
  url = {https://link.aps.org/doi/10.1103/PhysRevLett.87.220601}
}

@article{Lin,
  title = {Suppressing coherence effects in quantum-measurement-based engines},
  author = {Lin, Zhiyuan and Su, Shanhe and Chen, Jingyi and Chen, Jincan and Santos, Jonas F. G.},
  journal = {Phys. Rev. A},
  volume = {104},
  issue = {6},
  pages = {062210},
  numpages = {8},
  year = {2021},
  month = {Dec},
  publisher = {American Physical Society},
  doi = {10.1103/PhysRevA.104.062210},
  url = {https://link.aps.org/doi/10.1103/PhysRevA.104.062210}
}

@article{PhysRevA.108.062214,
  title = {Nonideal measurement heat engines},
  author = {Panda, Abhisek and Binder, Felix C. and Vinjanampathy, Sai},
  journal = {Phys. Rev. A},
  volume = {108},
  issue = {6},
  pages = {062214},
  numpages = {6},
  year = {2023},
  month = {Dec},
  publisher = {American Physical Society},
  doi = {10.1103/PhysRevA.108.062214},
  url = {https://link.aps.org/doi/10.1103/PhysRevA.108.062214}
}

@article{PhysRevE.98.042122,
  title = {Measurement-driven single temperature engine},
  author = {Ding, Xuehao and Yi, Juyeon and Kim, Yong Woon and Talkner, Peter},
  journal = {Phys. Rev. E},
  volume = {98},
  issue = {4},
  pages = {042122},
  numpages = {15},
  year = {2018},
  month = {Oct},
  publisher = {American Physical Society},
  doi = {10.1103/PhysRevE.98.042122},
  url = {https://link.aps.org/doi/10.1103/PhysRevE.98.042122}
}

@article{PhysRevE.107.054110,
  title = {Measurement-based quantum Otto engine with a two-spin system coupled by anisotropic interaction: Enhanced efficiency at finite times},
  author = {Purkait, Chayan and Biswas, Asoka},
  journal = {Phys. Rev. E},
  volume = {107},
  issue = {5},
  pages = {054110},
  numpages = {11},
  year = {2023},
  month = {May},
  publisher = {American Physical Society},
  doi = {10.1103/PhysRevE.107.054110},
  url = {https://link.aps.org/doi/10.1103/PhysRevE.107.054110}
}

@article{PhysRevE.95.032111,
  title = {Measurement-induced operation of two-ion quantum heat machines},
  author = {Chand, Suman and Biswas, Asoka},
  journal = {Phys. Rev. E},
  volume = {95},
  issue = {3},
  pages = {032111},
  numpages = {7},
  year = {2017},
  month = {Mar},
  publisher = {American Physical Society},
  doi = {10.1103/PhysRevE.95.032111},
  url = {https://link.aps.org/doi/10.1103/PhysRevE.95.032111}
}

@article{PhysRevE.103.032144,
  title = {Finite-time performance of a single-ion quantum Otto engine},
  author = {Chand, Suman and Dasgupta, Shubhrangshu and Biswas, Asoka},
  journal = {Phys. Rev. E},
  volume = {103},
  issue = {3},
  pages = {032144},
  numpages = {9},
  year = {2021},
  month = {Mar},
  publisher = {American Physical Society},
  doi = {10.1103/PhysRevE.103.032144},
  url = {https://link.aps.org/doi/10.1103/PhysRevE.103.032144}
}

@article{Sagawa2008,
  title = {Second Law of Thermodynamics with Discrete Quantum Feedback Control},
  author = {Sagawa, Takahiro and Ueda, Masahito},
  journal = {Phys. Rev. Lett.},
  volume = {100},
  issue = {8},
  pages = {080403},
  numpages = {4},
  year = {2008},
  month = {Feb},
  publisher = {American Physical Society},
  doi = {10.1103/PhysRevLett.100.080403},
  url = {https://link.aps.org/doi/10.1103/PhysRevLett.100.080403}
}

@article{groenewold1971,
  title={A problem of information gain by quantal measurements},
  author={Groenewold, Hilbrand J},
  journal={International Journal of Theoretical Physics},
  volume={4},
  number={5},
  pages={327--338},
  year={1971},
  publisher={Springer}
}

@article{Bresque_PhysRevLett.126.120605,
  title = {Two-Qubit Engine Fueled by Entanglement and Local Measurements},
  author = {Bresque, L\'ea and Camati, Patrice A. and Rogers, Spencer and Murch, Kater and Jordan, Andrew N. and Auff\`eves, Alexia},
  journal = {Phys. Rev. Lett.},
  volume = {126},
  issue = {12},
  pages = {120605},
  numpages = {6},
  year = {2021},
  month = {Mar},
  publisher = {American Physical Society},
  doi = {10.1103/PhysRevLett.126.120605},
  url = {https://link.aps.org/doi/10.1103/PhysRevLett.126.120605}
}

@article{Manikandan_PhysRevE.105.044137,
  title = {Efficiently fueling a quantum engine with incompatible measurements},
  author = {Manikandan, Sreenath K. and Elouard, Cyril and Murch, Kater W. and Auff\`eves, Alexia and Jordan, Andrew N.},
  journal = {Phys. Rev. E},
  volume = {105},
  issue = {4},
  pages = {044137},
  numpages = {10},
  year = {2022},
  month = {Apr},
  publisher = {American Physical Society},
  doi = {10.1103/PhysRevE.105.044137},
  url = {https://link.aps.org/doi/10.1103/PhysRevE.105.044137}
}

@article{Lisboa_PhysRevA.106.022436,
  title = {Experimental investigation of a quantum heat engine powered by generalized measurements},
  author = {Lisboa, V. F. and Dieguez, P. R. and Guimar\~aes, J. R. and Santos, J. F. G. and Serra, R. M.},
  journal = {Phys. Rev. A},
  volume = {106},
  issue = {2},
  pages = {022436},
  numpages = {8},
  year = {2022},
  month = {Aug},
  publisher = {American Physical Society},
  doi = {10.1103/PhysRevA.106.022436},
  url = {https://link.aps.org/doi/10.1103/PhysRevA.106.022436}
}

@article{Perna_PhysRevE.109.044102,
  title = {Limits on quantum measurement engines},
  author = {Perna, Guillermo and Calzetta, Esteban},
  journal = {Phys. Rev. E},
  volume = {109},
  issue = {4},
  pages = {044102},
  numpages = {11},
  year = {2024},
  month = {Apr},
  publisher = {American Physical Society},
  doi = {10.1103/PhysRevE.109.044102},
  url = {https://link.aps.org/doi/10.1103/PhysRevE.109.044102}
}

@article{Elouard_hpgc-nsmr,
  title = {Revealing the fuel of a quantum continuous measurement-based refrigerator},
  author = {Elouard, Cyril and Manikandan, Sreenath K. and Jordan, Andrew N. and Haack, G\'eraldine},
  journal = {Phys. Rev. E},
  volume = {113},
  issue = {1},
  pages = {014134},
  numpages = {16},
  year = {2026},
  month = {Jan},
  publisher = {American Physical Society},
  doi = {10.1103/hpgc-nsmr},
  url = {https://link.aps.org/doi/10.1103/hpgc-nsmr}
}

@article{Deffner_PhysRevE.94.010103,
  title = {Quantum work and the thermodynamic cost of quantum measurements},
  author = {Deffner, Sebastian and Paz, Juan Pablo and Zurek, Wojciech H.},
  journal = {Phys. Rev. E},
  volume = {94},
  issue = {1},
  pages = {010103(R)},
  numpages = {5},
  year = {2016},
  month = {Jul},
  publisher = {American Physical Society},
  doi = {10.1103/PhysRevE.94.010103},
  url = {https://link.aps.org/doi/10.1103/PhysRevE.94.010103}
}

@article{latune2025thermodynamically,
  title={A thermodynamically consistent approach to the energy costs of quantum measurements},
  author={Latune, Camille L and Elouard, Cyril},
  journal={Quantum},
  volume={9},
  pages={1614},
  year={2025},
  publisher={Verein zur F{\"o}rderung des Open Access Publizierens in den Quantenwissenschaften}
}

@article{guryanova2020ideal,
  title={Ideal projective measurements have infinite resource costs},
  author={Guryanova, Yelena and Friis, Nicolai and Huber, Marcus},
  journal={Quantum},
  volume={4},
  pages={222},
  year={2020},
  publisher={Verein zur F{\"o}rderung des Open Access Publizierens in den Quantenwissenschaften}
}

@article{Linpeng_PhysRevResearch.6.033045,
  title = {Quantum energetics of a noncommuting measurement},
  author = {Linpeng, Xiayu and Piccione, Nicol\`o and Maffei, Maria and Bresque, L\'ea and Prasad, Samyak P. and Jordan, Andrew N. and Auff\`eves, Alexia and Murch, Kater W.},
  journal = {Phys. Rev. Res.},
  volume = {6},
  issue = {3},
  pages = {033045},
  numpages = {8},
  year = {2024},
  month = {Jul},
  publisher = {American Physical Society},
  doi = {10.1103/PhysRevResearch.6.033045},
  url = {https://link.aps.org/doi/10.1103/PhysRevResearch.6.033045}
}

@article{Linpeng_PhysRevLett.128.220506,
  title = {Energetic Cost of Measurements Using Quantum, Coherent, and Thermal Light},
  author = {Linpeng, Xiayu and Bresque, L\'ea and Maffei, Maria and Jordan, Andrew N. and Auff\`eves, Alexia and Murch, Kater W.},
  journal = {Phys. Rev. Lett.},
  volume = {128},
  issue = {22},
  pages = {220506},
  numpages = {6},
  year = {2022},
  month = {Jun},
  publisher = {American Physical Society},
  doi = {10.1103/PhysRevLett.128.220506},
  url = {https://link.aps.org/doi/10.1103/PhysRevLett.128.220506}
}

@article{Baumgratz2014,
  title = {Quantifying Coherence},
  author = {Baumgratz, T. and Cramer, M. and Plenio, M. B.},
  journal = {Phys. Rev. Lett.},
  volume = {113},
  issue = {14},
  pages = {140401},
  numpages = {5},
  year = {2014},
  month = {Sep},
  publisher = {American Physical Society},
  doi = {10.1103/PhysRevLett.113.140401},
  url = {https://link.aps.org/doi/10.1103/PhysRevLett.113.140401}
}

@article{Tan,
  title = {Limitations of strong coupling in non-Markovian quantum thermometry},
  author = {Tan, Qing-Shou and Liu, Yang and Liu, Xulin and Chen, Hao and Xiao, Xing and Wu, Wei},
  journal = {Phys. Rev. A},
  volume = {112},
  issue = {4},
  pages = {042612},
  numpages = {13},
  year = {2025},
  month = {Oct},
  publisher = {American Physical Society},
  doi = {10.1103/t74h-c8kw},
  url = {https://link.aps.org/doi/10.1103/t74h-c8kw}
}

@article{PhysRevApplied.10.024003,
  title = {Phase-Tunable Thermal Logic: Computation with Heat},
  author = {Paolucci, Federico and Marchegiani, Giampiero and Strambini, Elia and Giazotto, Francesco},
  journal = {Phys. Rev. Appl.},
  volume = {10},
  issue = {2},
  pages = {024003},
  numpages = {10},
  year = {2018},
  month = {Aug},
  publisher = {American Physical Society},
  doi = {10.1103/PhysRevApplied.10.024003},
  url = {https://link.aps.org/doi/10.1103/PhysRevApplied.10.024003}
}

@article{ronzani2018tunable,
  title={Tunable photonic heat transport in a quantum heat valve},
  author={Ronzani, Alberto and Karimi, Bayan and Senior, Jorden and Chang, Yu-Cheng and Peltonen, Joonas T and Chen, ChiiDong and Pekola, Jukka P},
  journal={Nature Physics},
  volume={14},
  number={10},
  pages={991--995},
  year={2018},
  publisher={Nature Publishing Group UK London}
}

@article{lambert2023qutipbofin,
  title = {QuTiP-BoFiN: A bosonic and fermionic numerical hierarchical-equations-of-motion library with applications in light-harvesting, quantum control, and single-molecule electronics},
  author = {Lambert, Neill and Raheja, Tarun and Cross, Simon and Menczel, Paul and Ahmed, Shahnawaz and Pitchford, Alexander and Burgarth, Daniel and Nori, Franco},
  journal = {Phys. Rev. Res.},
  volume = {5},
  issue = {1},
  pages = {013181},
  numpages = {18},
  year = {2023},
  month = {Mar},
  publisher = {American Physical Society},
  doi = {10.1103/PhysRevResearch.5.013181},
  url = {https://link.aps.org/doi/10.1103/PhysRevResearch.5.013181}
}

@article{adesso2014continuous,
  title={Continuous variable quantum information: Gaussian states and beyond},
  author={Adesso, Gerardo and Ragy, Sammy and Lee, Antony R},
  journal={Open Syst. Inf. Dyn.},
  volume={21},
  number={01n02},
  pages={1440001},
  year={2014},
  publisher={World Scientific}
}
\bibliographystyle{ieeetr}  % Standard style for APS journals

\end{document}